\documentclass[nohyper,notoc]{JHEP3}
\usepackage{amsmath,amssymb,amsfonts,epsfig,dsfont}
\usepackage{latexsym}
\usepackage{graphicx}

\def\bse{\begin{subequations}}
\def\ese{\end{subequations}}

\newcommand{\Tr}{{\rm Tr\,}}

\newcommand{\D}{{\mathcal D}}

\newcommand{\beq}{\begin{equation}}
\newcommand{\be}{\begin{equation}}
\newcommand{\ee}{\end{equation}}
\newcommand{\bea}{\begin{eqnarray}}
\newcommand{\eea}{\end{eqnarray}}

\newcommand{\pa}{\partial}
\newcommand{\pab}{\bar{\partial}}

\newcommand{\nn}{\nonumber}
\newcommand{\zb}{\bar{z}}
\newcommand{\betab}{\bar{\beta}}

\newcommand{\Rb}{\bar{R}}

\newcommand{\Ab}{\bar{A}}

\newcommand{\kh}{\hat{k}}
\newcommand{\Eh}{\hat{E}}
\newcommand{\Et}{\tilde{E}}
\newcommand{\Ub}{\bar{U}}
\newcommand{\Phib}{\bar{\Phi}}
\newcommand{\phib}{\bar{\phi}}
\newcommand{\ub}{\bar{u}}

\newcommand{\etab}{\bar{\eta}}
\newcommand{\nub}{\bar{\nu}}
\newcommand{\Hh}{\hat{H}}

\newcommand{\uln}[1]{\underline{#1}}
\newcommand{\mc}{\mathcal}
\newcommand{\hb}{\bar{h}}
\newcommand{\fb}{\bar{f}}
\newcommand{\gb}{\bar{g}}
\newcommand{\Po}{{\mathcal P}}
\newcommand{\Ro}{{\mathcal R}}
\newcommand{\cE}{{\mathcal E}}
\newcommand{\mi}{\mid}
\newcommand{\ra}{\rangle}
\newcommand{\la}{\langle}

\title{General Leznov-Savelev solutions for Pohlmeyer reduced AdS$_5$ minimal surfaces.}

\author{Benjamin A. Burrington\\
Department of Physics, University of Toronto,\\ 60 St. George st, Toronto, ON M5S 1A7, Canada\\
{\tt E-mail: benburri@physics.utoronto.ca}
}

\abstract{We consider the Pohlmeyer reduced sigma model describing AdS$_5$ minimal surfaces.  We show that, similar to the affine Toda models, there exists a conformal extension to this model which admits a Lax formulation.  The Lax connection is shown to be valued in a ${\mathbb Z}_4$-invariant subalgebra of the affine Lie algebra $\widehat{su(4)}$.  Using this, we perform a modified version of a Laznov-Savelev analysis, which allows us to write formal expressions for the general solutions for the Pohlmeyer reduced AdS$_5$ theory.  This analysis relies on the a certain decomposition for the exponentiated algebra elements.}

\preprint{arXiv:xxxx.xxxx}


\begin{document}


\section{Introduction}


AdS/CFT has become an important tool in theoretical physics \cite{Maldacena:1997re} for understanding field theories at strong coupling, especially for theories with a large amount of symmetry.  Important among these are the superconformal theories dual to supergravity backgrounds, usually occurring as an AdS factor times a compact space \cite{Kachru:1998ys}.

In certain regimes, one can use classical minimal surfaces in AdS (or the compact space, or both) to calculate various quantities.  One can compute Wilson loop expectation values and quark/anti-quark potentials \cite{Maldacena:1998im,Drukker:2000rr,Berenstein:1998ij}, both of which are described by finding minimal surfaces that end on a prescribed smooth contour at the boundary of AdS.  Also by considering disconnected contours, one can compute Wilson loop-Wilson loop (or local operator) correlators \cite{Zarembo:1999bu,Drukker:2005cu}. One may also compute the the cusp anomalous dimension \cite{Gross:1998gk,Kruczenski:2002fb,Kruczenski:2007cy} by considering surfaces that end on continuous, but cusped contours at the boundary of AdS.  These quantities need to be regularized, and this is seen holographically by the minimal surfaces stretching to the boundary of AdS and having infinite surface areas.  Also, one can investigate aspects of large spin or large R-charge operators \cite{classicalStringSpin,Frolov:2003qc,Arutyunov:2003za,Arutyunov:2003uj}, which are dual to the ``spinning strings'' living in the bulk of AdS, and having possible profiles in the compact space as well.  String vertex operator correlators for an AdS target space have also been studied using minimal surfaces
\cite{Janik:2010gc,Buchbinder:2010vw,Roiban:2010fe,Ryang:2010bn,Hernandez:2010tg}, and so one may learn interesting features of string theories in non-trivial backgrounds.  Finally, one may also use these minimal surfaces to describe color ordered gluon scattering amplitudes \cite{GluonScat} using the dual piecewise lightlike Wilson loop.

There has been great progress made in the study of minimal surfaces in AdS due to the integrability of both the classical sigma model that describes them \cite{Arutyunov:2003za,Arutyunov:2003uj,Drukker:2005cu}, and the descendent Pohlmeyer reduced theories \cite{Pohlmeyer:1975nb,Pohlmeyer:1975nb2,Burrington:2009bh,Alday:2009dv}.  Integrability has also played an important role in uncovering certain aspects of the gauge theory duals as well.  For a good, although lengthy, review of recent advances in these topics, see \cite{Beisert:2010jr} and the accompanying articles.

Here we study minimal surfaces in AdS, focusing on the Pohlmeyer reduced
sigma model describing minimal surfaces in AdS$_5$.  The Pohlmeyer reduction can be thought of as a way of reducing the number of worldsheet fields by reducing to the physical modes.  This is done by specifying two directions of the spacetime using the worldsheet coordinates, although, somewhat implicitly.  The remaining modes are the ``gauge fixed'' degrees of freedom, although the gauge fixing has been accomplished in a more subtle way than a simple identification like ``$X^0(\tau,\sigma)=\tau$.''

For the symmetric cosets, like AdS, the classical sigma model is integrable, and in the AdS$_5$ case this has been show to descend to the reduced theory as well \cite{Burrington:2009bh,Alday:2009dv} in that they admit a zero curvature description, although, purely as a result of the reduction procedure itself.  Interestingly, the reduced theory for AdS$_3$ is a generalized sinh-Gordon model \cite{Alday:2009yn,Jevicki:2007aa}, and the AdS$_4$ case is a generalized affine-Toda (AT) model associated with the Lie algebra $so(5)$\footnote{we use lower case letters to denote Lie algebras, and capital letters to denote Lie groups}.  The AT models can be thought of generalizations to Toda models in the same way that the sinh-Gordon model is a generalization of the Liouville theory.  While in the AdS$_5$ case the Lax pair is valued in the Lie algebra $so(6)=su(4)$, it was shown in \cite{Burrington:2009bh,Alday:2009dv} that the AdS$_5$ case is not an AT model.  This can be most easily seen from the comments in \cite{Alday:2009dv}, where they note that the different components of the ``kinetic'' part of the Lax connection do not commute, which is in contrast to the affine-Toda models, where the kinetic parts are based on the Cartan subalgebra of the underlying Lie algebra.

While these models appear to have some degrees of freedom removed, they in fact contain almost all of the data needed to recover the minimal surfaces.  We explain this procedure for the AdS$_5$ case.  Given a solution to the Pohlmeyer reduced theory, one can construct the Lax connection.  Given this, one may use the six dimensional representation of $so(6)=su(4)$ and construct the Lax pair.  Solving this linear problem gives a solution for the $6$ homogeneous coordinates of AdS$_5$, hence furnishing a minimal area solution.

Next, we note that there is much that is known about Toda and related models.  For example, there exists a conformally extended version of the AT models, known as the conformal-affine-Toda (CAT) models \cite{Babelon:1990bq}.  Further, the AT models can be understood as conformally gauge fixed versions of the CAT models \cite{Constantinidis:1992hs}.  The CAT models have a Lax formulation, where the Lax connection is valued in an affine Lie algebra: the affine extension to the Lie algebra that the AT model is based on.  The affine extension contains two more Cartan generators, and likewise there are two additional field theoretic degrees of freedom.  In the AT model, the non-Cartan generators appearing in the Lax connection contain generators associated with $\pm$ the simple roots (we denote these $E_{\pm\alpha^{(i)}}$), as well as $\pm$ the longest root (we denote this $E_{\pm \theta}$).  When generalizing this to the CAT model, the generators $E_{\pm \theta}$ are promoted to generator associated with a simple root $E_{\pm \alpha^{(0)}}$.  This simple fact allows for the Leznov-Savelev (LS) analysis to be used, basically because the holomorphic indexed part of the lax connection (denoted $A$) has only positive ladder operators and Cartan generators.  The antiholomorphic indexed part of the Lax connection (denoted $\Ab$) similarly only has lowering ladder operators and Cartan generators.  This allows one to find a set of holomorphic and antiholomorphic functions that freely specify solutions (albeit formally), given that a certain decomposition is allowed.  The fact that only positive Ladder operators and Cartan generators are present in $A$ make highest weight states particularly important: these states ``ignore'' the raising ladder operators, and are simultaneous eigenvectors under the Cartan generators.

So, given this situation for the CAT models, one may wonder what parts of the CAT model constructions can be carried out for the Pohlmeyer reduced AdS$_5$ minimal surfaces.  First, the Pohlmeyer reduced AdS$_5$ model's Lax connection was shown to have a certain algebraic structure in \cite{Burrington:2009bh}, with the associated Lie algebra being $so(6)=su(4)$.  Clearly, one expects that there should be a conformally invariant model, given that the parent sigma model is conformally invariant.  This we address in section \ref{confextend} of our current work: we show that there exists a conformal extension with a Lax pair valued in the affine extension to $su(4)$ which we denote $\widehat{su(4)}$.  Further, we show that the ${\mathbb Z}_4$ symmetry found in \cite{Alday:2009dv} also has a good group theoretic meaning\footnote{it is a combination of rephasing by exponentiated Cartan generators and Weyl reflections}, and so can be extended to act on the base Lie algebra, and hence on the full affine Lie algebra.  In fact, we show that this ${\mathbb Z}_4$ (along with a rephasing using the grading operator $\D=-L_0$) defines a subalgebra which we denote $\hat{s}_4 \subset \widehat{su(4)}$.

One may then ask if there is a similar structure to the Lax pair, and so one might be able to carry out a LS-like analysis.  In fact this is true, however, one must contend with the fact that the kinetic terms of the connection are not diagonalizable, and so highest weight states are not what is important.  We find instead that instead the grade, defined for us to be the $\D=-L_0$ ``eigenvalue'' of the operators in the affine Lie algebra, is the important quantity (see the next subsection \ref{algprelim} for our conventions and notation for affine Lie algebras).  Hence, we will be concerned with highest grade states, rather than highest weight states.  These are simply the states related to the highest weight state by the ladder operators in the base Lie algebra: the 0 grade ladder operators.  We do indeed find a modified (LS) analysis that works for writing down general (formal) solutions in terms of holomorphic and antiholomorphic functions.  Further, the modification given by AdS$_3$ and AdS$_4$ models to the sinh-Gordon and AT $so(5)$ models can be seen to not affect the (LS) analysis (although this is quite trivial).  Again, the LS analysis is predicated on the existence of certain decompositions of the exponentiated elements of the algebra; the associated ``group.''

We give the following outline for the rest of the paper.  For the reader who wishes to skip the explanations in the text, we will give relevant equation numbers that will contain the bulk of the results.  We begin by offering a brief introduction to affine Lie algebras in the next subsection \ref{algprelim}.  The reader familiar with these may simply look at equation (\ref{comrel1}) and the enumerated list at the end \ref{algprelim}.  In section \ref{confextend} we present the conformally extended model.  We give two presentations of the model: the first is more easily connected to the usual presentation of CAT models, while the second is better suited for our purposes.  The results of this section needed for later are best presented in the Lax pair (\ref{Epnew})-(\ref{connectionnew}), or in terms of the equations of motion and Lagrangian (\ref{phi1eom})-(\ref{newaction}).  We discuss the symmetries of the Lax connection, and show that it is valued in a certain ${\mathbb Z}_4$-invariant subalgebra of the full $\widehat{su(4)}$ subalgebra, which we denote $\hat{s}_4$.  This restriction to a subalgebra, for our purposes, can be summarized in equations (\ref{gradeExpand}), (\ref{etildedef}) which gives the generator structures allowed for grades $-1,0,1$.  In section \ref{modLS} we present our modification to the LS proceedure, relying heavily on the fact that we are restricted to a subalgebra $\hat{s}_4$, and assuming that a certain decomposition of the exponentiated algebra is general.  The results of this analysis are best summarized by (\ref{cedef2})-(\ref{Mminusdiffeq2}) with the use of definitions (\ref{phidefnew}), (\ref{Lapmdef}), and setting $f_{\wedge}=\fb_{\wedge}=0$.  We end with a discussion of our results in section \ref{discussion}.  We give 3 appendices as well: appendix \ref{repsforsu4} contains the fundamental representations of $su(4)$ which we will need the ${\bf 4}$; appendix \ref{reductionviasymmetry} shows how to obtain the model of section \ref{confextend} via a symmetry reduction of the general $\widehat{su(4)}$ model; and finally appendix \ref{dimreduct} shows how to dimensionally reduce the model, giving an interesting one dimensional integrable system that, in principle, can furnish axisymmetric solutions that are minimal surfaces in AdS$_5$.

\subsection{Algebraic Preliminaries.}

\label{algprelim}

First we define the affine extension to the Lie algebra (see \cite{Kac:1990gs,DiFrancesco:1997nk}).  We start with the elements of a base Lie algebra $A, B,...$ along with the Lie bracket $[,]$.  We use the Lie bracket of this base algebra to define a new Lie bracket
\bea
&& [M\otimes t^m +\alpha \D  +a \kh,N\otimes t^n+\beta \D  + b \kh] \nn \\
&& \qquad \qquad =[M,N]\otimes t^{n+m} + \langle M,N\rangle  m \delta_{m,-n} \kh + \alpha n N \otimes t^n -\beta m M \otimes t^m \label{comrel1}
\eea
where $n,m$ are integers and $\langle ,\rangle $ is the Cartan-Killing form \footnote{we could normalize the term with $\kh$ on the right using any constant, but this simply comes down to redefining the normalization of $\kh$, or likewise using a different normalization for the Cartan-Killing form}.  We see that $\D $ counts the power of $t$, and $\kh$ is central.  In what follows, we will truncate notation to
\be
A\otimes t^n\equiv A^n.
\ee

We will need to define a notion of ``transposing'' the matrices in the affine case, and for this we will need certain facts about the base Lie algebra relevant for our disussion: $so(6)=su(4)$, the Lie algebra associated with the ${\rm A}_3={\rm D}_3$ root lattice.  This Lie algebra is generated by the Cartan subalgebra $H_1, H_2, H_3$ and a set of ladder operators $E_{[i,j,k]}$ where $i,j,k \in \{-1,0,1\}$ and $i^2+j^2+k^2=2$.  We will always use basis where $E_{[i,j,k]}^T=E_{[-i,-j,-k]},\; H_i^T=H_i$.  Further, it should be noted that the choice of normalization makes the Lie bracket simple (the structure constants are all $\pm 1$ or $0$), and we normalize the Cartan Killing form so that $\langle E_{\alpha},E_{-\alpha}\rangle =1$ for all $\alpha$, and no other terms and appear ($\langle E_{\alpha},E_{\beta}\rangle =0$ if $\beta\neq -\alpha$).  Further $\langle H_i,H_j\rangle =\delta_{i,j}$ as well.  Hence, we may write that $\langle A^T,B^T\rangle =\langle A,B\rangle $ for $so(6)=su(4)$, and our conventions match that of \cite{DiFrancesco:1997nk} given that we have $\alpha^2=2$ for all $\alpha$.  We note that $\D  =-L_0$ of \cite{DiFrancesco:1997nk}.  Now we construct a definition of ``transpose'' consistent with the notion that it should satisfy
\be
[{\mathcal M},{\mathcal N}]^T=[{\mathcal N}^T,{\mathcal M}^T]. \label{tranrelation}
\ee
We use the definition
\be
{\mathcal M}^T=\left(M\otimes t^m +\alpha \D  +a \kh\right)^T=M^T\otimes t^m -\alpha \D  -a \kh.
\ee
This relation is actually a map from the Lie algebra to itself, since we are taking that $M^T$ is still in the Lie algebra via $E_{[i,j,k]}^T=E_{[-i,-j,-k]},\; H_i^T=H_i$.  It is easy to check that the definition of $^T$ satisfies (\ref{tranrelation}), and further that the map ${\mathcal M}'=-{\mathcal M}^T$ defines an automorphism of the algebra\footnote{i.e. if $[{\mathcal M},{\mathcal N}]={\mathcal Q}$ then $[{\mathcal M}',{\mathcal N}']={\mathcal Q}'$.}.

Another useful operation that we will need later on is
\be
{\mathcal M}^O=\left(M\otimes t^m +\alpha \D  +a \kh\right)^O=M\otimes t^{-m} -\alpha \D  -a \kh.
\ee
One can see that
\be
[{\mathcal M},{\mathcal N}]^O=[{\mathcal M}^O,{\mathcal N}^O]. \label{orrelation}
\ee
We have denoted this operation $O$ because it denotes orientation reversing for the ``loop direction'' of the loop algebra, and exchanges ``winding'' for ``anti-winding.''

We take that the generators $E_{[i,j,k]}^m, H_i^n, \D , \kh$ are to be regarded as real, and so complex conjugation $^*$ works only on the coefficients.

Finally, we may also see that there is the automorphism of the algebra $\uln{H}_{1,3}=H_{1,3}, \uln{H}_2= -H_2$, $\uln{E}_{[i,j,k]}= E_{[i,-j,k]}$.  This is not generically a similarity transformation: it maps one representation to the complex conjugate representation.  We denote this by an underbar, reserving overbars to denote a $\zb$ index.  For the $6$ of $so(6)=su(4)$, this is a similarity transformation, but for the $4$, and other complex representations, it is not.  In fact, one can show that this map is the same as the map ${\mathcal M}'=-{\mathcal M}^T$ up to a similarity transformation.

Finally, to fix notation and naming conventions, we will note several terms that we will use throughout the rest of the paper.

\begin{enumerate}

\item $\D $: We will use the term ``grade'' to refer to the $\D $ eigenvalue.  This will be for both for the vector space on which representations work, as well as the generators themselves.  Hence, if a vector satisfied $\D  \mid \psi \rangle = m \mid \psi \rangle$, then $\mid \psi \rangle$ is said to have grade $m$.  A generator at grade $m$ is a generator ${\mathcal M}$ that satisfies $[\D ,{\mathcal M}]=m{\mathcal M}$, and so is necessarily of the form ${\mathcal M}=M\otimes t^m$.  This gives that the base Lie algebra along with $\D $ and $\kh$ are the grade 0 operators.  Further, for highest weight representations, the highest weight state will be fixed to have grade $0$ unless otherwise stated.  Thus, the vectors related by grade 0 operators to the highest weight state are all ``highest grade'' states, and fill out a single irreducible representation of the base Lie algebra.

\item Sometimes we will drop the $^0$ index on generators of the base Lie algebra.  Henceforth, any generators of the Lie algebra without a superscript are taken to be at grade 0.

\item $2H_1^0+H_2^0=2H_1+H_2$: We will use the term ``base level'' to refer to the level in the base Lie algebra of a given generator.  We will take the convention that $\alpha^{(1)}=[0,1,-1],\alpha^{(2)}=[1,-1,0],\alpha^{(3)}=[0,1,1]$ are the simple roots. In such a convention, the base level is measured by the generator $2H_1+H_2$.  Again, this terminology is to be applied to both the algebra elements and to the states of a highest weight representations, similar to the grade defined in item 1.

\item We recall the simple roots of the affine algebra are associated with the operators $E_{[-1,-1,0]}^1, E_{[0,1,-1]}^0,E_{[1,-1,0]}^0,E_{[0,1,1]}^0$, where we have given them in $\alpha_0,\alpha_1, \alpha_2, \alpha_3$ order.  Given a simple Lie algebra $a$, we will refer to the affine extension as $\hat{a}$.  Hence, the algebra important for us is the $\widehat{su(4)}$ algebra.

\item We will refer to the ``ladder level'' in what follows.  This is given by the total number of simple ladder operators needed to make a given ladder operator.  By definition, the Cartan subalgebra has ladder level 0.  The ladder level is measured by $4\D  + 2H_1^0+H_2^0$.  This guarantees that the ladder level of the generators associated with the simple roots $E_{[-1,-1,0]}^1, E_{[0,1,-1]}^0,E_{[1,-1,0]}^0,E_{[0,1,1]}^0$ all have ladder level 1.  Again, we may apply the terminology to to both states and generators, as in items 1 and 3.
\end{enumerate}


\section{The conformally extended model.}

\label{confextend}

\subsection{The conformally extended model: first presentation.}

Our first task is to show that the Pohlmeyer reduced sigma model describing minimal surfaces in AdS$_5$ admits a conformal extension.  This is not too surprising given that the original sigma model possess this symmetry.  The question is how to implement this, while retaining integrability, and this is answered by looking at Pohlmeyer reduced sigma models for AdS$_3$ and AdS$_4$.  These are both affine-Toda (AT) models, and the process for conformally extending these models, given by the conformal-affine-Toda (CAT) models, was presented in \cite{Babelon:1990bq}, and a number of interesting studies \cite{CATstudies} (see also \cite{Aratyn:1990tr} for some related models).

In the AT model, the generator associated with minus the longest root, $E_{-\theta}$, appears in the holomorphic indexed component of the Lax connection ($A_z=A$), summed with the simple roots: $e_+=E_{-\theta}+\sum_{\alpha \mbox{ simple}} E_{\alpha}$.  In the CAT model this is promoted to $e_+'=E_{-\theta}^1+\sum_{\alpha \mbox{ simple}} E_{\alpha}^0$: in the affine Lie algebra this is the sum of the simple ladder operators.  Similar comments hold for the antiholomorphic indexed component ($A_{\zb}=\Ab$) of the Lax connection, although using instead $e_+^{T O}$.  In the Pohlmeyer reduced AdS$_5$ theory, the same set of generators $E_{-\theta}+\sum_{\alpha \mbox{ simple}} E_{\alpha}$ appears \cite{Burrington:2009bh}, and so one is tempted to generalize the connection in the same way.  This intuition is almost correct, and we define instead
\bea
E_+&=&U(z)\left(E^0_{[1,-1,0]}+E^2_{[-1,-1,0]}\right) + E^0_{[0,1,-1]}+E^0_{[0,1,1]} \label{Ep}\\
E_-&=&\Ub(\zb)\left(E^0_{[-1,1,0]}+E^{-2}_{[1,1,0]}\right) + E^0_{[0,-1,1]}+E^0_{[0,-1,-1]} \label{Em}\\
\Eh&=&E^{-1}_{[1,0,-1]}-E^1_{[-1,0,1]} + E^{-1}_{[1,0,1]}-E^1_{[-1,0,-1]} \label{Eh}
\eea
and we further define
\be
\phi=\phi_1 H_1^0+\phi_2 H_2^0 +\eta \D  + (\delta_1 H_1^0 + \delta_2 H_2^0 +\delta_3 H_3^0)\eta + \nu \kh. \label{defphiorig}
\ee
The important point is to note is that $[E_+,\Eh]=[E_-,\Eh]=0$, mimicking the same behavior of the Pohlmeyer reduced AdS$_5$ case, and further indicating why the generator $E_{-\theta}$ is promoted to $E_{-\theta}^2$ rather than $E_{-\theta}^1$.  This is done to avoid half integer $\D $ grades being assigned to the generators in $\Eh$.  We also recall that $\phi_1$ is to be regarded as pure imaginary to make contact with Euclidean AdS$_5$ minimal surfaces \cite{Burrington:2009bh}.

The Lax connection for this enlarged system is given by
\bea
A&=&-\pa \phi + \frac{\lambda}{\sqrt{2}}e^{{\rm ad}_\phi}E_+ +\frac12 \frac{\pa \Lambda_1}{\cosh^2(\phi_1)}e^{-{\rm ad}_\phi} \Eh \nn \\
\Ab&=& \pab \phi + \frac{1}{\lambda \sqrt{2}}e^{-{\rm ad}_\phi}E_- +\frac12 \frac{\pab \Lambda_1}{\cosh^2(\phi_1)}e^{{\rm ad}_\phi} \Eh
\eea
and the Lax pair is simply
\bea
&&\pa \Psi(z,\zb;\lambda)=A(z,\zb;\lambda) \Psi \\
&&\pab \Psi(z,\zb;\lambda)=\Ab(z,\zb;\lambda) \Psi.
\eea
We further take
\be
\delta_1=1, \quad \delta_2=\frac 12, \quad \delta_3=0
\ee
in (\ref{defphiorig}).  The choice for $\delta_3$ and $\delta_1$ is found necessary by the integrability condition for the Lax connection written above.  The choice of $\delta_2$, however, is simply a convenience, and could be absorbed into shifts of the definitions of $\phi_2$.  However, it has the virtue of making the coefficient of $\eta$ in $\phi$ a combination of the grade and base level operators, $\D$ and $2H_1+H_2$.  Also note that the generator multiplying $\eta$ in $\phi$ ($H_1^0+\frac12 H_2^0+\D $) commutes with $\Eh$:
\be
\left[H_1^0+\frac12H_2^0+\D ,\Eh\right]=0.
\ee
Therefore, the action of ${\rm ad}_\phi$ on $\Eh$ gives no contributions of
$\eta$ (similarly for $[\pa \phi,\Eh]$).

The integrability condition for the Lax pair is (assuming $\Psi$ is invertible)
\be
\pab A- \pa \Ab +[A,\Ab]=0.
\ee
Writing out this field strength, we find
\bea
\pab A- \pa \Ab +[A,\Ab]=&&-2\pa\pab \phi + \frac12 \pab\left(\frac{\pa \Lambda_1}{\cosh^2(\phi_1)}\right)e^{-{\rm ad}_{\phi}}\Eh - \frac12 \pa\left(\frac{\pab \Lambda_1}{\cosh^2(\phi_1)}\right)e^{{\rm ad}_{\phi}}\Eh \nn \\
&& - \frac{\pa \Lambda_1}{\cosh^2(\phi_1)}\left[\pab \phi, e^{-{\rm ad}_{\phi}}\Eh\right]- \frac{\pab \Lambda_1}{\cosh^2(\phi_1)}\left[\pa \phi, e^{{\rm ad}_{\phi}}\Eh\right] \\
&&+\frac12 \left[e^{{\rm ad}_{\phi}}E_+,e^{-{\rm ad}_{\phi}}E_-\right]+\frac14\frac{\pa \Lambda_1\pab \Lambda_1}{\cosh^4(\phi_1)}\left[e^{-{\rm ad}_{\phi}}\Eh,e^{{\rm ad}_{\phi}}\Eh\right].\nn
\eea
One may evaluate this easily because all commutation relations in $su(4)$ are implied by the explicit representation of appendix \ref{repsforsu4}, which in turn imply the commutation relations in $\widehat{su(4)}$.  Using this we may write out the equations of motion for the fields
\bea
&&-2\pa\pab \phi_1 +\frac{\pa \Lambda_1 \pab \Lambda_1}{\cosh^4(\phi_1)}\sinh(2\phi_1)+U\Ub e^{-2\phi_2}\sinh(2\phi_1)e^{\eta}=0 \\
&&-2\pa\pab \phi_2 +\left(e^{2\phi_2}-U\Ub\cosh(2\phi_1)e^{-2\phi_2}\right)e^{\eta}=0 \\
&&-2\pa \pab \eta=0 \\
&&-2\pa \pab \nu - \sinh(2\phi_1)\frac{\pa\Lambda_1 \pab \Lambda_1}{\cosh^4(\phi_1)} +  U \Ub e^{-2\phi_1-2\phi_2+\eta}=0 \\
&&\frac12 \pab \left(\frac{\pa \Lambda_1}{\cosh^2(\phi_1)}\right)e^{\phi_1}-\frac12 \pab \left(\frac{\pab \Lambda_1}{\cosh^2(\phi_1)}\right)e^{-\phi_1} \nn \\
&&\qquad \qquad-\frac{\pa \Lambda_1}{\cosh^2(\phi_1)}e^{-\phi_1}\pab\phi_1-\frac{\pab \Lambda_1}{\cosh^2(\phi_1)}e^{\phi_1}\pa\phi_1 \nn \\
&&= \frac{-1}{2\sinh(\phi_1)}\left(\pa\left(\tanh^2(\phi_1)\pab \Lambda_1\right)+\pab\left(\tanh^2(\phi_1)\pa \Lambda_1\right)\right)=0
\eea
where above we have already substituted in $\pa\pab \eta=0$ into other equations of motion.
From these expressions, a Lagrangian formulation is easily found to be
\bea
{\mathcal{L}}=&&\pa \phi_1 \pab \phi_1 + \pa \phi_2 \pab \phi_2 + \tanh^2(\phi_1)\pa \Lambda_1 \pab \Lambda_1 + \frac12\left(U(z) \Ub(z) e^{-2\phi_2}\cosh(2\phi_1)+e^{2\phi_2}\right)e^\eta \nn \\
&&+ (\pa \nu \pab \eta + \pab \nu \pa \eta)+ \frac12\left(\pa \eta (2\pab \phi_1+ \pab \phi_2) +\pab \eta (2\pa \phi_1+ \pa \phi_2)\right).
\eea
%

\subsection{The conformally extended model: second, and preferred, presentation.}

The reader not interested in the details of the transformation connecting the last presentation to the new presentation may simply proceed to equation (\ref{Epnew})-(\ref{connectionnew}) for the new Lax pair, and (\ref{phi1eom})-(\ref{newaction}) for the new equations of motion, which are just a field redefinition of the above.

The above Lax pair is not the most symmetric presentation one can concoct.  We have presented it in the above way to make contact with previous work on the conformal affine Toda model.  One obtains the CAT model for the affine Lie algebra $\widehat{so(5)}$ by setting $\Lambda_1={\rm constant}$:   $\widehat{so(5)}$ has been embedded into $\widehat{so(6)}=\widehat{su(4)}$ in the natural way except that we are restricted to the even grade operators.  Restricting the affine algebra to even grade is finding an $\widehat{so(6)}$ subalgebra inside of $\widehat{so(6)}$; one then restricts to an $\widehat{so(5)}$ subalgebra of this $\widehat{so(6)}$ subalgebra to find the CAT model associated with $\widehat{so(5)}$.

To motivate the following transformations, we simply note that the coefficient of $\eta$ appearing in $\phi$ is $\D +(2H_1+H_2)/2$.  This operator treats all of the generators of $E_{+}$ ($E_{-}$) identically, and we find $[\D +(2H_1+H_2)/2,E_{\pm}]=\pm1/2 E_{\pm}$.  Further $[\D +(2H_1+H_2)/2,\Eh]=0$, suggesting certain shifts in the definition of the grading operator.

Keeping the above comment in mind, we perform some manipulations of the algebra to make the presentation more symmetric.  First, we note that we may redefine certain generators.  For example, we may define the new generators
\be
\Hh_1^0=H_1^0+2\times \kappa \kh, \qquad \Hh_2^0=H_2^0+\kappa \kh, \qquad \Hh_3^0=H_3^0.
\ee
Note that again we will be taking advantage of the combination ``$2H_1+H_2$,'' explaining the relative factor of 2 between the shifts by $\kh$.  Now, we note that
\be
[\Hh_i^0,[\mathcal{M}]]=[H_i^0,[\mathcal{M}]]
\ee
because $\kh$ is central, and so the new and old $H_i^0$ generators are equivalent when appearing in commutators.  However, they are different when they appear as a result of a commutator.  Note that the only time that $H_i^0$ shows up on the right hand side is when we are dealing with a commutator of the following form
\be
\left[E_{[i,j,k]}^{m},E_{[-i,-j,-k]}^{-m}\right]=iH_1^0+jH_2^0+kH_3^0+m\langle E_{[i,j,k]},E_{[-i,-j,-k]}\rangle \kh.\label{formh}
\ee
which we may rewrite
\bea
&& \left[E_{[i,j,k]}^{m},E_{[-i,-j,-k]}^{-m}\right] \\
&& = i\Hh_1^0+j\Hh_2^0+k\Hh_3^0+\left(-\frac{\left[2\kappa i+\kappa j\right]}{\langle E_{[i,j,k]},E_{[-i,-j,-k]}\rangle }+m\right)\langle E_{[i,j,k]},E_{[-i,-j,-k]}\rangle \kh. \nn
\eea
Above, the appearance of the Cartan-Killing form in the denominator may be ignored because $\langle E_{[i,j,k]},E_{[-i,-j,-k]}\rangle=1$, and so we find
\bea
&& \left[E_{[i,j,k]}^{m},E_{[-i,-j,-k]}^{-m}\right] \nn \\
&& = i\Hh_1^0+j\Hh_2^0+k\Hh_3^0+\left(-\left[2\kappa i+\kappa j\right]+m\right)\langle E_{[i,j,k]},E_{[-i,-j,-k]}\rangle \kh \nn \\
&& = i\Hh_1^0+j\Hh_2^0+k\Hh_3^0+\left(\kappa\langle -[(2H_1+H_2), E_{[i,j,k]}],E_{[-i,-j,-k]}\rangle +m\langle E_{[i,j,k]},E_{[-i,-j,-k]}\rangle\right) \kh. \nn \\
\eea
In fact, one may rewrite the entire algebra in terms of the above rules finding that
\bea
&& [M\otimes t^m +\alpha \D  +a \kh,N\otimes t^n+\beta \D  + b \kh] \nn \\
&& \qquad =[M,N]\otimes t^{n+m} + (-\kappa \langle [2H_1+H_2,M],N\rangle + m\langle M,N\rangle ) \delta_{m,-n} \kh \nn \\
&& \qquad \qquad + \alpha n N \otimes t^n -\beta m M \otimes t^m
\eea
because the only place that the new term appears is in commutators of the form already explored (\ref{formh}).  This is because if $M$ and $N$ are cartan generators, then the commutator vanishes (they are base level 0), and in cases of missmatched ladder operators, the Cartan-Killing form returns 0.  This is obviously just some rewriting of the affine Lie algebra we started with.


One may redefine more generators as well.  We will be concerned with the shift
\be
\hat{\D}=\D + \kappa'\left(2H_1^0+H_2^0\right).
\ee
Now, because $\D $ does not appear as the result of any commutator, and we leave the definitions of $H_1^0$ and $H_2^0$ fixed, we must simply see what the effect is when $\hat{\D}$ appears in a commutator.  Let us consider a general generator of the base Lie algebra $M_p$ of base level $p$.  We find that
\be
[\hat{\D},M_p^m]=(m+\kappa' p) M_p^m
\ee
and we define a new index for such operators $m'=m+\kappa' p$.  The algebra becomes
\be
[M_p^{m'},N_q^{n'}]=[M_p,N_q]^{m'+n'}+(-\kappa p \langle M_p,N_q\rangle + (m'-\kappa' p)\langle M_p,N_q\rangle ) \delta_{m'-\kappa'p,-n'+\kappa'q} \kh.
\ee
In the first part of the above equality, the condensed notation $[M_p,N_q]^{m'+n'}$ means to evaluate $[M_p,N_q]$ in the base Lie algebra, and then add the tensor product with $t$ to the power given by ${m'+n'}$.  This works exactly because levels add in the commutator: if $M_p$ is level $p$ and $N_q$ is level $q$, then $[M_p,N_q]$ is level $p+q$.  Further, the inner products $\langle M_p,N_q\rangle $ are only non zero when $p=-q$, and so $\delta_{m'-\kappa'p,-n'+\kappa'q}=\delta_{m',-n'}$ when multiplying this term.  So, now modifying the algebra a second time, we find
\bea
&& [M\otimes t^{m'} +\alpha \hat{\D} +a \kh,N\otimes t^{n'}+\beta \hat{\D} + b \kh] \nn \\
&& \qquad =[M,N]\otimes t^{n'+m'} + ((-\kappa-\kappa') \langle [2H_1+H_2,M],N\rangle + m'\langle M,N\rangle ) \delta_{m',-n'} \kh \nn \\
&& \qquad \qquad + \alpha n' N \otimes t^{n'} -\beta m' M \otimes t^{m'}.
\eea
We can immediately see from the above discussion that the relevant factor for us is $\kappa'=\frac12$, so that the $\hat{\D}$ is the generator multiplying $\eta$ in the original presentation of the model.  Then, to make the algebra as simple as possible, we take $\kappa=-\frac12$.  We then rescale $\hat{\D}$ by $2$ and scale $\eta$ by $\frac12$ so that we get integer grade associated with all of the generators and further, we must rescale the $\kh$ operator by $1/2$ so that the coefficient appearing as the result of a commutator is an integer.  We then shift the definition of $\nu$ to absorb the factors of $\phi_i$ that come from the redefinitions of the $H_i^0$, and further rescale this $\nu$ by $2$ to absorb the rescaling of $\kh$ to make the Lax connection look simple.  We will then drop all of the notation introduced to denote the shifted generators and fields.

This all has a very simple effect.  First, we define the new generators
\bea
E_+&=&U(z)\left(E^{1}_{[1,-1,0]}+E^{1}_{[-1,-1,0]}\right) + E^{1}_{[0,1,-1]}+E^{1}_{[0,1,1]} \label{Epnew}\\
E_-&=&\Ub(\zb)\left(E^{-1}_{[-1,1,0]}+E^{-1}_{[1,1,0]}\right) + E^{-1}_{[0,-1,1]}+E^{-1}_{[0,-1,-1]} \label{Emnew}\\
\Eh&=&E^{0}_{[1,0,-1]}-E^0_{[-1,0,1]} + E^{0}_{[1,0,1]}-E^0_{[-1,0,-1]} \label{Ehnew}
\eea
again satisfying $[E_+,\Eh]=[E_-,\Eh]=0$, and which again are related to the old generators simply by the new labeling scheme.  Further, after the above transformations, the field $\phi$ becomes
\be
\phi=\phi_1 H_1^0+\phi_2 H_2^0 +\eta \D  + \nu \kh. \label{phidefnew}
\ee
and the new Lax connection for our system is given exactly as before
\bea
A&=&-\pa \phi + \frac{\lambda}{\sqrt{2}}e^{{\rm ad}_\phi}E_+ +\frac12 \frac{\pa \Lambda_1}{\cosh^2(\phi_1)}e^{-{\rm ad}_\phi} \Eh \nn \\
\Ab&=& \pab \phi + \frac{1}{\lambda \sqrt{2}}e^{-{\rm ad}_\phi}E_- +\frac12 \frac{\pab \Lambda_1}{\cosh^2(\phi_1)}e^{{\rm ad}_\phi} \Eh. \label{connectionnew}
\eea
We stress here that the redefinition of $\D , m$ and $H_i^0, \kh$ has resulted in {\it no change} to the definition of the algebra, and so we find
\bea
&& [M\otimes t^m +\alpha \D  +a \kh,N\otimes t^n+\beta \D  + b \kh] \nn \\
&& \qquad \qquad =[M,N]\otimes t^{n+m} + \langle M,N\rangle  m \delta_{m,-n} \kh + \alpha n N \otimes t^n -\beta m M \otimes t^m.
\eea
In this way, the redefinitions can almost be thought of as an automorphism of the algebra: however, we should be careful to note that we have again redefined the grade indices to be integer valued by scaling some operators.  A final note is in order: above, the spectral parameter $\lambda$ appears always as $\lambda^n E_{\alpha}^n$, and so the power of $\lambda$ matches the grade of the operator it multiplies.  Hence, it may be gauged completely away.  Any change to the phase or scale of $\lambda$ may be accomplished by conjugation with $\exp(i\theta \D )\exp(\rho \D )$.  Hence any transformation of $\lambda$ may actually be accounted for in a ``group theoretic'' way using $\D $.

Next, one should note that we have absorbed the factors of $\kh$ that came from shifting the $H_i^0$ by changing the definition of $\nu$ above: this will change its equation of motion, but by a simple field redefinition.  There was also the rescaling of $\eta$ above that changes the $\eta$ equation of motion as well, but again by a simple field redefinition.

For completeness, we write down the equations of motion, and find
\bea
&&-2\pa\pab \phi_1 +\frac{\pa \Lambda_1 \pab \Lambda_1}{\cosh^4(\phi_1)}\sinh(2\phi_1)+U\Ub e^{-2\phi_2}\sinh(2\phi_1)e^{2\eta}=0   \label{phi1eom}\\
&&-2\pa\pab \phi_2 +\left(e^{2\phi_2}-U\Ub\cosh(2\phi_1)e^{-2\phi_2}\right)e^{2\eta}=0 \label{phi2eom}\\
&&-2\pa \pab \eta=0 \label{etaeom}\\
&&-2\pa \pab \nu + U \Ub e^{-2\phi_2+2\eta} \cosh(2\phi_1)+e^{2\phi_2 +2\eta}=0 \label{nueom}\\
&&\frac12 \pab \left(\frac{\pa \Lambda_1}{\cosh^2(\phi_1)}\right)e^{\phi_1}-\frac12 \pab \left(\frac{\pab \Lambda_1}{\cosh^2(\phi_1)}\right)e^{-\phi_1} \nn \\
&&\qquad \qquad-\frac{\pa \Lambda_1}{\cosh^2(\phi_1)}e^{-\phi_1}\pab\phi_1-\frac{\pab \Lambda_1}{\cosh^2(\phi_1)}e^{\phi_1}\pa\phi_1 \nn \\
&&= \frac{-1}{2\sinh(\phi_1)}\left(\pa\left(\tanh^2(\phi_1)\pab \Lambda_1\right)+\pab\left(\tanh^2(\phi_1)\pa \Lambda_1\right)\right)=0 \label{Lambda1eom}
\eea
which follows from the action
\bea
{\mathcal{L}}=&&\pa \phi_1 \pab \phi_1 + \pa \phi_2 \pab \phi_2 + \tanh^2(\phi_1)\pa \Lambda_1 \pab \Lambda_1 + \frac12\left(U(z) \Ub(z) e^{-2\phi_2}\cosh(2\phi_1)+e^{2\phi_2}\right)e^{2\eta} \nn \\
&&+ (\pa \nu \pab \eta + \pab \nu \pa \eta). \label{newaction}
\eea
It is clear that the above action and equations of motion are simply a field redefinition from the last section, obtained by shifting $\nu$ and then rescaling $\nu$ and $\eta$, and it is now obvious exactly how to do this.

In either presentation of the model, it is clear that $U$ and $\bar{U}$ can be absorbed into a shift of $\eta$ and $\phi_2$.  This has the interesting effect that models with different $U$ are related simply by different solutions to the same equations.

Further, after removing $U$ and $\Ub$, we see that the remaining model is conformally invariant.  Under the transformation
\be
z=f(z'), \qquad \zb=\bar{f}(\zb')
\ee
the fields map in a simple way
\bea
&& \phi_i(z,\zb)=\phi_i'(z',\zb'), \quad  \Lambda_1(z,\zb)=\Lambda_1'(z',\zb')\nn \\
&& \eta(z,\zb)=\eta'(z',\zb')-\frac{1}{2}\left(\ln(\pa_{z'} f)+\ln(\pa_{\zb'} \bar{f})\right), \\
&& \nu(z,zb)\rightarrow \nu'(z',\zb') - \frac{c}{2}\left(\ln(\pa_{z'} f)-\ln(\pa_{\zb'} \bar{f})\right) \nn
\eea
where $c$ is arbitrary.

Finally, we may relate the above conformal symmetry to the original conformal symmetry of the sigma model.  We may shift $\phi_2$ by $\phi_2=\phi_2'-\eta$.  Plugging this in to the action with $U$s removed (and then removing the prime from the new $\phi_2$), we find
\bea
{\mathcal{L}}=&&\pa \phi_1 \pab \phi_1 + \pa \phi_2 \pab \phi_2 + \tanh^2(\phi_1)\pa \Lambda_1 \pab \Lambda_1 + \frac12\left(e^{4\eta}e^{-2\phi_2}\cosh(2\phi_1)+e^{2\phi_2}\right) \nn \\
&&+ (\pa \nu \pab \eta + \pab \nu \pa \eta)-(\pa\phi_2\pab\eta+\pa\eta\pab\phi_2)+\pa\eta \pab \eta.
\eea
In this presentation, $e^{4\eta}$ plays the role of $U\Ub$.  The modified conformal symmetry is easy to read as
\bea
&& \phi_i(z,\zb)=\phi_i'(z',\zb')-\delta_{i2}\frac{1}{2}\left(\ln(\pa_{z'} f)+\ln(\pa_{\zb'} \bar{f})\right), \nn \\
&& \Lambda_1(z,\zb)=\Lambda_1'(z',\zb')\nn \\
&& \eta(z,\zb)=\eta'(z',\zb')-\frac{1}{2}\left(\ln(\pa_{z'} f)+\ln(\pa_{\zb'} \bar{f})\right), \\
&& \nu(z,zb)\rightarrow \nu'(z',\zb') - \frac{c}{2}\left(\ln(\pa_{z'} f)-\ln(\pa_{\zb'} \bar{f})\right). \nn
\eea
Note that this says that $e^{2\phi_2}$ is weight $(1,1)$.  This is exactly the assignment coming from the original identification in the sigma model (using our earlier conventions \cite{Burrington:2009bh}) $\pa Y \cdot \pab Y\equiv e^{2\phi_2}$ where $Y^I$ are the homogeneous coordinates of AdS$_5$.  Further, we see that $e^{8\eta}$ is weight $(4,4)$.  We recall the definitions of \cite{Burrington:2009bh} $\pa^2 Y \cdot \pa^2 Y \equiv U^2$, and $\pab^2 Y \cdot \pab^2 Y \equiv \Ub^2$.  Hence, we find that $e^{8\eta}$ is playing the role of $U^2 \Ub^2=\pa^2 Y \cdot \pa^2 Y \pab^2 Y \cdot \pab^2 Y$ which is indeed a $(4,4)$ worldsheet field.

\subsection{Symmetries of lax connection and a subalgebra.}

We have now converted the above Lax pair into a suitable form.  We shall now explore some of its interesting transformation properties.  First we construct a $Z_4$ transformation already found in \cite{Alday:2009dv}.  We will use the explicit $4\times 4$ representation of the matrices given in the appendix to explore it's action on the base Lie algebra.  First, define the matrix
\be
C=\begin{pmatrix} 0 & 0 & 0 & \frac{-1+i}{\sqrt{2}} \\ 0 & 0 & \frac{1+i}{\sqrt{2}} & 0 \\ 0 & \frac{1-i}{\sqrt{2}} & 0 & 0 \\ \frac{-1-i}{\sqrt{2}} & 0 & 0 & 0 \end{pmatrix}.
\ee
We will use the matrix $C$ to perform a similarity transformation of the base Lie algebra.  We will use a loose notation where we will denote the element $C^{-1}M C\otimes t^m= C^{-1}M^m C$: this, by definition, has no action on $\D $ or $\kh$.  Recall that the transformation ${\mathcal M}^T$ has been defined so as to fit the relation (\ref{tranrelation}): because we have mapped the algebra into itself, we use the same definition on the new generators.  However, this transformation is specific to the ${\bf 4}$ representation of the base Lie algebra.  Highest weight representations of the affine Lie algebra always contain different representations of the base Lie algebra at each grade.  Hence, we must come up with some group theoretic meaning to this transformation, and we do this below, showing that this ${\mathbb Z}_4$ action can always be constructed for any representation, and has a specific group theoretic meaning.

First, we explore the action of $C$ (along with the transpose) on the base lie algebra, and see that this acts as the replacements
\bea
&& C^{-1}H_i^TC= -H_i \;\; {\mbox{ $i=1,2$}}, \quad C^{-1}H_3^TC= H_3, \quad C^{-1}(e^{{\rm ad}_\phi}E_+)^TC = i e^{{\rm ad}_\phi}E_+,\nn \\
&& C^{-1}(e^{-{\rm ad}_\phi}E_{-})^TC = -i e^{-{\rm ad}_\phi}E_{-},\quad C^{-1}(e^{\pm{\rm ad}_\phi}\Eh)^T C= -e^{\pm{\rm ad}_\phi}\Eh \label{cmapping}
\eea
where we have simply dropped the superscripts to project the generators into the base Lie algebra.

Next, to help identify $C$, we note another $Z_4$ action given by conjugation by
\be
C_h = e^{\frac{-i\pi}{2}(2H_1+H_2)}.
\ee
This mapping can in fact be used on the entire affine Lie algebra because it is well defined for all representations. Using $C_h$, we find the mapping acts as
\bea
C_h^{-1}H_iC_h= H_i {\mbox{ $i=1,2,3$}}, \quad C_h^{-1}\D C_h= \D , \quad C_h^{-1}\kh C_h= \kh \\
C_h^{-1}E_+C_h = i E_+,\quad  C_h^{-1}E_{-}C_h = -i E_{-},\quad C_h^{-1}\Eh C_h= -\Eh
\eea
(we may include or ignore the ${\rm ad}_\phi$ because the $H_i$ commute with these generators).  This is easy to check because $2H_1+H_2$ is the level operator.  The base level of the matrices in $E_+$ are all $1$ or $-3$ which are congruent modulo $4$.  Similarly, $E_-=E_{+}^T$ is constructed using the transposed matrices: i.e. those that are base level $-1$ or $3$.  Finally, $\Eh$ is constructed using matrices that are base level $2$ or $-2$, which are again congruent modulo 4.  Clearly $C_h$ works this way for any representation of the Lie algebra, and so it's action is well defined on the entire affine Lie algebra as well.

We are now in a position to identify $C$.  We find that $C=C_{W_{12}} C_h$ where $C_{W_{12}}$ is a matrix that performs a Weyl reflection: it takes $E_{[i,j,k]}\rightarrow E_{[-i,-j,k]}$.  This is accomplished by reflecting the through the plane orthogonal to $[1,-1,0]$ first (taking $[i,j,k]\rightarrow [j,i,k]$) and then reflecting through the plane orthogonal to $[1,1,0]$ (taking $[i,j,k]\rightarrow [-j,-i,k]$).  We accomplish this mapping by a similarity transformation, $C_{W_{12}}$, and the must also take $H_1\rightarrow - H_1, H_2\rightarrow -H_2, H_3\rightarrow H_3$.  This essentially undoes the transpose for the generators $E_{i,j,k}$ because nowhere does $H_3$ appear in the ${\rm ad}_\phi$ action, which means that the matrices $E_{[i,j,k]}$ and $E_{[i,j,-k]}$ get dressed with the same exponential factors of the fields.  This translates into the fact that $E_{[i,j,k]}+E_{[i,j,-k]}\rightarrow_{_{T}} E_{[-i,-j,-k]}+E_{[-i,-j,k]}\rightarrow_{_{C_{W_{12}}}} E_{[i,j,-k]}+E_{[i,j,k]}$.  This then clarifies the structure of this ${\mathbb Z}_4$ symmetry generated by $C$: it is the combination of a Weyl reflection and the $Z_4$ generated by $C_h$.  This also shows that such a $Z_4$ is present using any representation, because any Weyl reflection can always be implemented as a similarity transformation, and $C_h$ is defined in any representation.  Hence, the transformation
\bea
A'= C^{-1}A^T C,\quad \Ab'=C^{-1}\Ab^T C
\eea
in fact makes sense in any representation.  This means that these transformations make sense at the level of the base Lie group, or as an automorphism of the algebra, and so can be defined on the entire affine Lie algebra as an action of the base Lie group.  Another way of thinking about this is that $C_h$ may be trivially extended to act on the full affine Lie algebra because it is a base group element, and the base Weyl reflections are a subset of the Weyl reflections available in the affine Lie algebra.  Thus, we can define an action of the base Lie algebra on the whole affine Lie algebra, along with the action of $^T$ as
\bea
&& C^{-1}H_i^TC= -H_i {\mbox{ $i=1,2$}}, \quad C^{-1}H_3^TC= H_3, \quad C^{-1}\D ^TC=-\D , \quad C^{-1}\kh^TC=-\kh, \nn \\
&& C^{-1}(e^{{\rm ad}_\phi}E_+)^TC = i e^{{\rm ad}_\phi}E_+, \label{cmappingtot} \\
&&C^{-1}(e^{-{\rm ad}_\phi}E_{-})^TC = -i e^{-{\rm ad}_\phi}E_{-},\quad C^{-1}(e^{\pm{\rm ad}_\phi}\Eh)^T C= -e^{\pm{\rm ad}_\phi}\Eh \nn
\eea
where above all the generators are understood to be in the full affine Lie algebra.

There are other generators one may wish to consider as well, for example
\be
C_U(\theta)=e^{-i(\D )\theta}
\ee
for any $\theta$.  This acts as
\bea
&& C_U(\theta)^{-1}H_iC_U(\theta)= H_i, \quad C_U(\theta)^{-1}\D C_U(\theta)= \D , \quad C_U(\theta)^{-1}\kh C_U(\theta)= \kh \\
&& C_U(\theta)^{-1}E_+C_U(\theta) = e^{i\theta} E_+,\quad  C_U(\theta)^{-1}E_{-}C_U(\theta) = e^{-i\theta} E_{-},\quad C_U(\theta)^{-1}\Eh C_U(\theta)= \Eh \nn
\eea
(again ignoring the adjoint action because $\D $ commutes with $\phi$) using the presentation (\ref{Epnew}), (\ref{Emnew}), (\ref{Ehnew}) and so acts as a continuous $U(1)$ symmetry shifting the phase of the spectral parameter $\lambda$.  This can be used to account for the extra factors of $i$ appearing in the ${\mathbb Z}_4$ appearing above.

Therefore, we may now combine these two symmetries to find that the Lax connection obeys
\bea
P(A(\lambda))\equiv C^{-1}\left(-\left[C_U(\pi/2)^{-1}A(\lambda)C_U(\pi/2)\right]^T\right)C=A(\lambda), \qquad \nn \\ P(\Ab(\lambda))\equiv C^{-1}\left(-\left[C_U(\pi/2)^{-1}\Ab(\lambda)C_U(\pi/2)\right]^T\right)C=\Ab(\lambda).
\eea
This is now a purely algebraic statement: it in fact defines a subalgebra of the full $\widehat{su(4)}$ algebra.  This is easy to see because if $M$ and $N$ are members of the affine lie algebra satisfying $P(M)=M, P(N)=N$ then if $Q=[M,N]$, then
\be
Q=[M,N]= [P(M),P(N)]=P(Q)
\ee
and so $Q=P(Q)$ for any resultant commutator of the restricted elements $P(M)=M$: the restriction closes under the Lie bracket and defines a subalgebra.  This fact will prove crucial in the Leznov-Saveliev analysis in the next section.

One may, in addition to the above, require that
\be
C^{-1}_U(\theta) A\left(e^{-i\theta}\lambda\right) C_U(\theta)=A\left(\lambda\right). \label{u1restriction}
\ee
This is a $U(1)$ that guarantees that the grade of an operator also denotes the power of the spectral parameter that multiplies it.  Given these two symmetries, the general $\widehat{su(4)}$ Lax connection can be reduced to the form  (\ref{Epnew})-(\ref{connectionnew}), (along with a reality condition) which we do in the appendix \ref{reductionviasymmetry}.

In what follows, it will be important to know the general generator structure allowed at each grade, given the restriction $P(M)=M$.  For the first few grades, this is given by
\bea
&& \mbox{grade $0$}\sim \left[H_1^0, H_2^0, \D , \kh, \Eh, \Et \right] \nn \\
&& \mbox{grade $1$}\sim \left[ E_{[1,-1,0]}^{1}, E_{[0,1,-1]}^{1}+ E_{[0,1,1]}^{1}, E_{[-1,-1,0]}^{1},E_{[0,-1,-1]}^1-E_{[0,-1,1]}^1\right]  \label{gradeExpand} \\
&& \mbox{grade $-1$}\sim \left[ E_{[-1,1,0]}^{-1}, E_{[0,-1,1]}^{-1}+ E_{[0,-1,-1]}^{-1}, E_{[1,1,0]}^{-1},E_{[0,1,1]}^{-1}-E_{[0,1,-1]}^{-1}\right].\nn
\eea
where the terms in brackets are the various operators allowed at the prescribed grade.  We have defined a new convenient generator structure
\be
\Et=E^{0}_{[1,0,-1]}+E^0_{[-1,0,1]} + E^{0}_{[1,0,1]}+E^0_{[-1,0,-1]} \label{etildedef}
\ee
which will become important in the next section.  Further, for future reference, we will name this algebra
\be
\hat{s}_4=\left\{A \in \widehat{su(4)}| P(A)=A\right\}.
\ee
%


\section{Modified Leznov-Saveliev analysis}


\label{modLS}

We are now in a position to consider a Leznov-Saveliev (LS) analysis \cite{Leznov:1979td}, see also \cite{Babelon} for a complete discussion of this technique applied to the conformal-Toda (CT) models (given by the AT models by removing the generators associated with $\pm \theta$, the longest root).  The results of this section will be somewhat more speculative because they depend on the existence of a ``group'' associated to the affine Lie algebra, quite similar to the CAT case \cite{Babelon:1990bq} (see also \cite{Aratyn:1990tr}).  It may seem that we have depended on these in the last section, however, we have mainly used the group elements as a crutch to produce certain effects at the level of the algebra, and so the results of the last section depend only on automorphisms of the algebra.  Further, the LS analysis will depend on some decompositions of the ``group'' being available, which we will see as we proceed.

We stress that we will be using the presentation (\ref{Epnew})- (\ref{connectionnew}), resulting in equation of motion (\ref{phi1eom})-(\ref{Lambda1eom}) henceforth.

Before proceeding to the LS analysis, we need to make one comment about the structure of the differential equation we are trying to solve.  The equation of motion for $\Lambda_1$ is given by a conservation law
\be
\pa (\tanh^2(\phi_1) \pab \Lambda_1)+\pab (\tanh^2(\phi_1) \pa \Lambda_1)=0.
\ee
In two dimensions, conservation laws are equivalent to the vanishing of a $U(1)$ field strength.  This means that the above may be written in terms of a ``pure gauge'' condition, i.e. it is equivalent to the local existence of a scalar function $\Lambda_2$ satisfying
\be
\tanh^2(\phi_1)\pa \Lambda_1 =\pa \Lambda_2, \quad \tanh^2(\phi_1) \pab \Lambda_1 = -\pab \Lambda_2.
\ee

\subsection{The form of the solution.}

Now we turn to the full LS analysis.  The equations of motion are given by the zero curvature condition
\be
\pab A - \pa \Ab + [A,\Ab]=0,
\ee
and so the solution must be pure gauge
\be
A=-T^{-1} \pa T, \quad \Ab=-T^{-1}\pab T.
\ee
In the usual LS analysis, one introduces a decomposition based on $\phi$ in the next step.  However, recall that we have more interesting kinetic terms which need to be treated on similar footing.  Therefore, we introduce the algebra elements
\be
\Lambda_+=\frac{\Lambda_1+\Lambda_2}{2}\Eh, \quad \Lambda_-=\frac{\Lambda_1-\Lambda_2}{2}\Eh. \label{Lapmdef}
\ee
Now, given this, we consider a decomposition of $T$ as
\be
T=g_1e^{-\Lambda_-}e^{\phi},
\ee
which is completely general, given that as of now $g_1$ is a general exponentiated algebra element.
This gives that
\bea
-T^{-1}\pa T&=&-\pa\phi+ e^{-\phi}\pa \Lambda_-e^{\phi}-e^{-\phi}e^{\Lambda_-}g_1^{-1}\pa g_1 e^{-\Lambda_1}e^{\phi}\nn \\
&=& -\pa \phi+ \frac{\pa\Lambda_1}{2\cosh^2(\phi_1)}e^{-{\rm ad}_\phi}\Eh-e^{-\phi}e^{\Lambda_-}g_1^{-1}\pa g_1 e^{-\Lambda_1}e^{\phi}
\eea
where we have used $\pa \Lambda_2=\tanh^2(\phi_1)\pa \Lambda_1$ to eliminate $\pa \Lambda_2$.  Note that the first two terms agree with the derivative terms of $A$ appearing in (\ref{connectionnew}).  Hence, equating with $A$ we find that
\be
-g_1^{-1}\pa g_1=\frac{1}{\sqrt{2}} e^{-\Lambda_-}e^{2\phi}E_+ e^{-2\phi} e^{\Lambda_-}.\label{g1eq}
\ee
Note that $E_+$ is positive grade, and $\phi$ and $\Lambda_-$ are grade 0.  Therefore, the operator appearing on the right hand side of the above equation is zero when applied to any highest grade state.  This is the first indication of how to modify the LS analysis for the model at hand: we will be concerned with the {\it grade} in the algebra, rather than the ladder level.  Hence we see that
\bea
\pa(g_1 \mid \mu \rangle) &=& (\pa g_1) \mid \mu \rangle \nn \\
&=& -g_1 \frac{1}{\sqrt{2}} e^{-\Lambda_-}e^{2\phi}E_+ e^{-2\phi} e^{\Lambda_-}\mid \mu \rangle=0,
\eea
for $\mid \mu \rangle$ a constant highest grade state.  Therefore $g_1\mid \mu \rangle$ is an antiholomorphic vector.

Now we will comment ont the decomposition that we need to assume.  We note that the $\hat{s}_4$ algebra can be decomposed into three subalgebras, depending on the grade of the operators.  We will call these ${\mathcal N}_-$ for those operators with grade less than 0, ${\mathcal N_0}$ to be those operators with grade 0, ${\mathcal N}_+$ to be those operators with grade greater than 0: clearly each of these are subalgebras because the grade adds in the commutator.  It is also clear that
\be
\hat{s}_4={\mathcal N}_-\oplus {\mathcal N}_0 \oplus {\mathcal N}_+,
\ee
which descends to the universal enveloping algebra in a particular way.  We assume that the above decomposition allows for an arbitrary group element $g$ to be decomposed as
\be
g=N_0 N_- N_+ \label{decompform}
\ee
where $N_i\in \exp({\mathcal N_i})$.  Note that if one could prove such a theorem, one could in principle put the factors in any order, as there is no real distinction between positive and negative grade\footnote{It is convention of the sign of the $\D$ operator.  Conversely, one can argue that given one ordering (\ref{decompform}), then the other form must also be available if one asserts that the inverse of an arbitrary ``group element'' is also an arbitrary group element. The inverse reverses the order of the $N_i$, but then the $N_0$ may be brought though, replacing $N_\pm=\exp(n_{\pm})$ by $N_\pm=\exp(N_0^{-1} n_{\pm}N_0)$ (with $n_\pm \in {\mathcal N}_{\pm}$).  It is clear that $N_0^{-1} n_{\pm}N_0\in {\mathcal N}_{\pm}$.}.  We note that to move and commuting $N_0$ around is trivial because $N_0^{-1}{\mathcal N}_{\pm}N_0={\mathcal N}_{\pm}$, though the exact representative from this subalgebra generically changes.  Hence, we assume that every group element $g$ admits the decompositions
\be
g=N_0 N_- N_+= N_0' N_+' N_-'
\ee
as well as others where the $N_0$ term appears in any order.  The usual NS analysis relies on an analogous decomposition, only using the ladder level as the separation between the various ${\mathcal N}_i$.  There is an obvious connection to this and the original presentation of the model, restricted to even grade, and so removing the $\Lambda_1$ term from the discussion.

We now proceed further, and take the decomposition
\be
g_1=M_- N_+
\ee
where $N_+$ is a member of $\exp({\mathcal N}_+)$, and $M_-$ is the exponential of negative and zero grade operators (the first part of the decomposition (\ref{decompform})).  It is clear that $N_+\mid \mu \rangle=\mid \mu \rangle$ if $\mid \mu \rangle$ is highest grade.  One may immediatly guess that $M_-$ is antiholomorphic, which one can show by expanding out (\ref{g1eq}) as
\be
-M_-^{-1}\pa M_-=\pa N_+ N_+^{-1} +\frac{1}{\sqrt{2}}N_+ e^{-\Lambda_-}e^{2\phi}E_+ e^{-2\phi} e^{\Lambda_-}N_+^{-1}
\ee
and noting that the left hand side is made of operators at grade 0 or less, while the right hand side is made of operators at grade 1 or more.  Therefore, we conclude that
\be
-M_-^{-1}\pa M_-=0
\ee
and so $M_-$ is antiholomorphic.  One may similarly set $-T^{-1}\pab T=\Ab$ using the above decompositions and show that
\bea
-M_-^{-1}\pab M_-&=&\pab N_+ N_+^{-1}+2 N_+ e^{-\Lambda_-}\pa \phi e^{\Lambda_-} N_+^{-1} \nn \\
&& +\frac{(1-\tanh^2(\phi_1))}{2}N_+ e^{-\Lambda_-}e^{2\phi}\Eh e^{-2\phi} e^{\Lambda_-} N_+^{-1} \nn \\
&& -\frac{(1+\tanh^2(\phi_1))}{2}N_+ \Eh N_+^{-1} +\frac{1}{\sqrt{2}}N_+^{-1} E_- N_+ \label{Mminusmess}
\eea
where in writing the above we have made use of the fact that $[E_+,\Eh]=0\rightarrow [E_+, \Lambda_\pm]=0$.  Equation (\ref{Mminusmess}) is difficult to read except for one interesting feature: the right hand side has operators with grade -1 or greater.  The grade -1 operator is simply $(1/\sqrt{2})E_-$, and so we may read that
\bea
-M_-^{-1}\pab M_-&=&\pab \hb_1(\zb) H_1^0 + \pab \hb_2(\zb)H_2^0 + \pab\gb_{\D}(\zb)\D+\pab \gb_{\kh}(\zb)\kh\nn \\
 && + \fb_{\wedge}(\zb)\Eh+\fb(\zb)\Et+\frac{1}{\sqrt{2}} E_- \label{Mminusdiffeq}
\eea
where we have simply written out the arbitrary algebra element at grade 0 in $\hat{s}_4$ with arbitrary holomorphic functions.  The partial derivatives appearing on functions are for later convenience.  Hence, the solution of the above differential equation is a path ordered exponential, however, it is a one dimensional path ordered exponential.  To simplify, one can define
\bea
&& \bar{Q}_{00}=\exp(-\hb_i H_i^0 -\gb_{\D} \D - \gb_{\kh} \kh) \nn \\
&& \bar{Q}_{01}=\Po\exp\left(\int d\zb \bar{Q}_{00}[\fb_{\wedge}(\zb)\Eh+\fb(\zb)\Et]\bar{Q}_{00}^{-1}\right)  \\
&& \bar{Q}_-=\Po\exp\left(\int d\zb \bar{Q}_{01}\bar{Q}_{00}\frac{1}{\sqrt{2}}E_-\bar{Q}_{00}^{-1}\bar{Q}_{01}^{-1}\right) \nn
\eea
with $\Po$ denoting path ordering.  This gives one way of denoting the solution as
\be
M_-=\bar{Q}_-\bar{Q}_{01}\bar{Q}_{00} \label{qbarprod}
\ee
which is useful in that it explicitly gives the dependence on the functions multiplying the Cartan subalgebra of $\widehat{su(4)}$.  We will use this to identify the coefficient $g_\D$ later.

Now we go back to the original element $T$, and instead define a new decomposition
\be
T=g_2 e^{\Lambda_+}e^{-\phi}.
\ee
using this decomposition, one similarly finds that
\be
\pab g_2^{-1} g_2=\frac{1}{\sqrt{2}}e^{\Lambda_+}e^{-2\phi}E_- e^{2\phi} e^{\Lambda_+}.
\ee
This gives that $\langle \mu \mid g_2^{-1}$ is a holomorphic vector, given that $\langle \mu \mid$ is the Hermitian conjugate of a highest grade state (i.e. all negative grade operators operating to the left on $\langle \mu \mid$ annihilate this state).  Again, we assume a decomposition, however now of the form
\be
g_2=M_+ N_-
\ee
with $N_-\in \exp({\mathcal N}_-)$ and $M_+$ is is the exponential of positive and zero grade operators.  Using the same steps as above, we find that
\be
-M_+^{-1}\pab M_+=\pab N_- N_-^{-1} + \frac{1}{\sqrt{2}} N_- e^{-\Lambda_+}e^{-2\phi}E_-e^{2\phi}e^{\Lambda_+}N_-^{-1}
\ee
and so comparing grades we see that  $M_+^{-1}\pab M_+=0$ and so $M_+$ is holomorphic.  Also, completely analogously we see that
\bea
\pa M_+^{-1}M_+&=&-M_+^{-1}\pa M_{+}=\pa N_- N_-^{-1}-2N_-e^{-\Lambda_+}\pa\phi e^{\Lambda_+} N_-^{-1} \nn \\
&&+\frac{1-\tanh^2(\phi_1)}{2}\pa\Lambda_1 N_- e^{-\Lambda_+}e^{-2\phi} \Eh e^{2\phi} e^{\Lambda_+}N_-^{-1} \nn \\
&&-\frac{1+\tanh^2(\phi_1)}{2}\pa\Lambda_1 N_- \Eh N_-^{-1}+\frac{1}{\sqrt{2}}N_- E_+ N_-^{-1}.
\eea
Again, by examining the grade of each operator, we see that the above equation implies that
\bea
\pa M_+^{-1} M_+&=&\pa h_1(z)H_1+\pa h_2(z)H_2+\pa g_{\D}(z)\D+\pa g_{\kh}\kh \nn \\
&& + f_{\wedge}(z) \Eh + f(z) \Et+\frac{1}{\sqrt{2}} E_+. \label{Mplusdiffeq}
\eea
Again, we may take this and write
\bea
&& {Q}_{00}=\exp(h_i H_i +g_{\D} \D + g_{\kh} \kh) \nn \\
&& {Q}_{01}=\Ro \exp\left(\int d\zb \bar{Q}_{00}^{-1}[f_{\wedge}(\zb)\Eh+f(\zb)\Et]\bar{Q}_{00}\right)  \\
&& {Q}_+=\Ro \exp\left(\int d\zb \bar{Q}_{01}^{-1}\bar{Q}_{00}^{-1}\frac{1}{\sqrt{2}}E_-\bar{Q}_{00}\bar{Q}_{01}\right) \nn
\eea
(where $\Ro$ denotes reverse path ordering) so that we may write
\be
M_+^{-1}=Q_{00}Q_{01}Q_+. \label{qprod}
\ee

Finally, we see that by requiring $T^{-1}T=1$ that we can find
\be
g_2^{-1}g_1=e^{-\Lambda_+}e^{-2\phi}e^{\Lambda_-}\equiv \cE.
\ee
Rewriting, we see that
\be
N_-^{-1} M_+^{-1} M_- N_+ =e^{-\Lambda_+}e^{-2\phi}e^{\Lambda_-}\equiv \cE \label{cedef}.
\ee
Again, we can see the assumed decomposition in the above: $M_+^{-1} M_-$ is some general group element, which we assume can be decomposed as $M_+^{-1} M_-= N_- N_0 N_+^{-1}$, and so $N_+$ and $N_-$ simply play the role of the matrices necessary to bring $M_+^{-1} M_-$ to an exponentiated grade 0 operator.

We may already identify some of the degrees of freedom using (\ref{cedef}).  First, note that $\pa \pab \eta=0$ and so we decompose $\eta$ as
\be
2\eta=\eta_+(z)+\eta_-(\zb).
\ee
We may take (\ref{cedef}) and take it's expectation value between two highest grade vectors (with non zero $\D$ eigenvalues), in which case
\be
\langle \mu' \mid M_+^{-1} M_- \mid \mu \rangle>=\langle \mu' \mid e^{-\Lambda_+}e^{-2\phi}e^{\Lambda_-} \mid \mu \rangle.
\ee
Hence, on the left there are exponentials of $\D$ coming from $Q_{00}$ and $\bar{Q}_{00}$ using expressions (\ref{qbarprod}) and (\ref{qprod}).  Since $\D$ may not be produced by commutators, these exponential factors may simply be equated.  Thus, we find that
\be
\eta_-(\zb)=\gb_\D(\zb),\quad \eta_+(z)=-g_{\D}(z).
\ee

\subsection{Extracting the remaining functions.}

We now will proceed to write down the form of the solutions.  Note that we have the set of fields $\phi_1, \phi_2, \Lambda_1, \Lambda_2$ and $\nu$ left to find, while the degrees of freedom associated with $\eta$ have already been accounted for.  It will prove sufficient to consider only one irreducible representation of the $\widehat{su(4)}$ algebra, namely the representation with 4 highest grade states, in Dynkin notation the $(1,0,0,0)$ representation, where the first 3 entries are the weights that are dual to the simple roots of the base Lie algebra.  Restricting to the highest grade states, the generators of the form $M\otimes t^0$ are given by a ${\bf 4}$ representation when acting on these states.  Henceforth, we shall also assume that the $\D$ eigenvalue of these highest grade states is 0 (this is merely a convention).  This is in fact a fundamental representation, and it is required to have $\kh$ eigenvalue $1$ because all of the comarks $su(4)$ are all 1 (see \cite{DiFrancesco:1997nk}).

We therefore find it convenient to explore the ${\bf 4}$ representation of the $su(4)$ Lie algebra, which we give explicitly in appendix \ref{repsforsu4}.  In this representation, we find that
\be
\Eh= \begin{pmatrix} 0 & 0 & -1 & 0 \\ 0 & 0 & 0 & -1 \\ 1 & 0 & 0 & 0 \\ 0 & 1 & 0 & 0\end{pmatrix}.
\ee
We will find it convenient to use eigenvectors of $\Eh$ in this representation
\bea
&& \mid 1 \rangle = \frac{1}{\sqrt{2}}\begin{pmatrix} 1 \\ 0 \\ -i \\ 0\end{pmatrix}, \qquad \mid 2 \rangle = \frac{1}{\sqrt{2}}\begin{pmatrix} 0 \\ 1 \\ 0 \\ -i\end{pmatrix} \nn \\
&& \mid 3 \rangle = \frac{1}{\sqrt{2}}\begin{pmatrix} 1 \\ 0 \\ i \\ 0\end{pmatrix}, \qquad \mid 4 \rangle = \frac{1}{\sqrt{2}}\begin{pmatrix} 0 \\ 1 \\ 0 \\ i\end{pmatrix}
\eea
satisfying
\bea
&& \Eh \mid 1 \rangle = i \mid 1 \rangle, \quad \Eh \mid 2 \rangle = i \mid 2 \rangle, \quad \Eh \mi 3 \ra =-i \mi 3 \ra, \quad \Eh \mid 4 \ra = -i \mid 4 \ra \nn \\
&& \la 1 \mi \Eh = \la 1 \mi i , \quad \la 2 \mi \Eh  = \la 2 \mi i , \quad \la 3 \mi \Eh =\la 3 \mi (-i) , \quad \la 4 \mi \Eh =\la 1 \mi (-i)
\eea
where as usual $\la k \mi = (\mi k \ra)^{\dagger}$.  Other operators that we will need include $\Et$:
\bea
&& \Et \mid 1 \rangle = i \mid 3 \rangle, \quad \Et \mid 2 \rangle = i \mid 4 \rangle, \quad \Et \mi 3 \ra =-i \mi 1 \ra, \quad \Et \mid 4 \ra = -i \mid 2 \ra \nn \\
&& \la 1 \mi \Et = \la 3 \mi (-i) , \quad \la 2 \mi \Et  = \la 4 \mi (-i) , \quad \la 3 \mi \Et =\la 1 \mi (i) , \quad \la 4 \mi \Et =\la 2 \mi (i)
\eea
and $H_1$
\bea
&& H_1 \mid 1 \rangle = 1/2 \mid 3 \rangle, \quad H_1 \mid 2 \rangle = 1/2 \mid 4 \rangle, \quad H_1 \mi 3 \ra =1/2 \mi 1 \ra, \quad H_1 \mid 4 \ra = 1/2 \mid 2 \ra \nn \\
&& \la 1 \mi H_1 = \la 3 \mi 1/2 , \quad \la 2 \mi H_1  = \la 4 \mi 1/2 , \quad \la 3 \mi H_1 =\la 1 \mi 1/2 , \quad \la 4 \mi H_1 =\la 2 \mi 1/2 \nn \\
\eea
and $H_2$
\bea
&& H_2 \mid 1 \rangle = 1/2 \mid 1 \rangle, \quad H_2 \mid 2 \rangle = -1/2 \mid 2 \rangle, \quad H_2 \mi 3 \ra =1/2 \mi 3\ra, \quad H_2 \mid 4 \ra = -1/2 \mid 4 \ra \nn \\
&& \la 1 \mi H_2 = \la 1 \mi 1/2 , \quad \la 2 \mi H_2  = \la 2 \mi (-1/2) , \quad \la 3 \mi H_2 =\la 3 \mi 1/2 , \quad \la 4 \mi H_2 =\la 4 \mi (-1/2). \nn \\
\eea

Now, given the above relations in the base lie algebra, it is clear how to extend this to the full affine Lie algebra.  We will use the same notation as the above to denote the highest grade states in the $(1,0,0,0)$ representation, however, we have the additional information
\be
\kh \mi i \ra=\mi i \ra, \quad \la i \mid \kh= \la i \mi, \quad \D \mi i \ra =0, \la i \mi \D=0,
\ee
and of course the ladder operators $E_{\alpha}^n\mid i \rangle=0$ for $n\geq 1$.  The lowering operators with negative grade are well defined, but we will not have to use these in what follows.  Hence, we know a certain subspace of states in the affine Lie algebra, namely the highest grade states of the $(1,0,0,0)$ representation.

Next, we introduce a notation
\be
\langle i \mid M_+^{-1} M_- \mid j \rangle = \langle i \mid \cE \mid j \rangle=\langle i \mid e^{-\Lambda_+}e^{-2\phi}e^{\Lambda_-} \mid j \rangle \equiv \langle i j\rangle.
\ee
Note that the left hand most expression is the easiest to take derivatives of, given that $M_+^{-1}$ is holomorphic and satisfies the differential equation given in (\ref{Mplusdiffeq}), and $M_-$ is antiholomorphic, and satisfies the differential equation (\ref{Mminusdiffeq}).

We may quickly find all of the fields using the above notation
\bea
\la 1 1 \ra =e^{-i\Lambda_2}e^{-2\nu}e^{-\phi_2}\cosh(\phi_1), \quad \la 1 3 \ra =-e^{-i\Lambda_1}e^{-2\nu}e^{-\phi_2}\sinh(\phi_1) \nn \\
\la 3 1 \ra =-e^{i\Lambda_1}e^{-2\nu}e^{-\phi_2}\sinh(\phi_1), \quad \la 3 3 \ra =e^{i\Lambda_2}e^{-2\nu}e^{-\phi_2}\cosh(\phi_1) \nn \\
\la 22 \ra =e^{-i\Lambda_2}e^{-2\nu}e^{\phi_2}\cosh(\phi_1), \quad \la 2 4 \ra =-e^{-i\Lambda_1}e^{-2\nu}e^{\phi_2}\sinh(\phi_1) \nn \\
\la 4 2 \ra =-e^{i\Lambda_1}e^{-2\nu}e^{\phi_2}\sinh(\phi_1), \quad \la 4 4 \ra =e^{i\Lambda_2}e^{-2\nu}e^{-\phi_2}\cosh(\phi_1).
\eea
Any inner products mixing odd and even numbers vanish, i.e. $\la 1 2\ra=\la 1 4\ra=\la 2 1\ra=\la 2 3\ra=\la 4 1\ra=\la 4 3\ra=0$.  This is because of the ${\mathbb Z_4}$ restriction on the affine Lie algebra, which in turn restricts the grade 0 operators.  We note that of the $8$ possible $\la i j\ra$ there are $5$ independent combinations of these, which we give as
\bea
&& e^{2 i \Lambda_2}=\frac{\la 33 \ra}{\la 11 \ra}, \quad e^{2i\Lambda_1}=\frac{\la 31 \ra}{\la 13 \ra}, \nn \\
&& e^{2\phi_2}=\frac{\la 2 4 \ra}{\la 13 \ra}, \quad \tanh^2(\phi_1)=\frac{\la 13 \ra \la 3 1 \ra}{\la 11\ra \la33 \ra} \\
&& e^{-4\nu}=\la 22 \ra \la 33 \ra - \la 31 \ra \la 24 \ra.
\eea
In addition, there 3 relations
\be
\frac{\la 33\ra }{\la 11 \ra}=\frac{\la 44 \ra}{\la 22 \ra}, \quad \frac{\la 11\ra }{\la 13 \ra}=\frac{\la 22 \ra}{\la 24 \ra}, \quad \frac{\la 24\ra }{\la 13 \ra}=\frac{\la 42 \ra}{\la 31 \ra}. \label{redundant1}
\ee
Derivable from the above relations, one can also show
\be
\frac{\la 11\ra }{\la 31 \ra}=\frac{\la 22 \ra}{\la 42 \ra}, \quad \frac{\la 33\ra }{\la 13 \ra}=\frac{\la 44 \ra}{\la 24 \ra}, \quad \frac{\la 44\ra }{\la 42 \ra}=\frac{\la 33\ra}{\la 31 \ra} \label{redundant2}
\ee
which we will use later.  Hence, all of the functions are determined entirely in terms of the holomorphic and antiholomorphic functions in $M_+^{-1} M_-$.

One should note that we have six holomorphic functions $h_i, g_{\D}, g_{\kh}, f_{\wedge}, f$ and six antiholomorphic counterparts.  This is one degree of freedom too many for the five dynamical fields $\phi_i, \eta, \nu, \Lambda_1$.  However, in the next section, we will show that one holomorphic and one antiholomorphic function must be set to zero.

\subsection{Checking the solution.}

We now will check the solutions we have written down.  Therefore, we assume the form of the solution to be
\be
N_-^{-1} M_+^{-1} M_- N_+ =e^{-\Lambda_+}e^{-2\phi}e^{\Lambda_-}\equiv \cE \label{cedef2}.
\ee
with $M_+^{-1}$ determined by
\bea
\pa M_+^{-1} M_+&=&\pa h_1(z)H_1+\pa h_2(z)H_2-\pa \eta_{+}(z)\D+\pa g_{\kh}\kh \nn \\
&& + f_{\wedge}(z) \Eh + f_{\sim} \Et+\frac{1}{\sqrt{2}} E_+. \label{Mplusdiffeq2}
\eea
and $M_-$ determined by
\bea
-M_-^{-1}\pab M_-&=&\pab \hb_1(\zb) H_1^0 + \pab \hb_2(\zb)H_2^0 + \pab\eta_{-}(\zb)\D+\pab \gb_{\kh}(\zb)\kh\nn \\
 && + \fb_{\wedge}(\zb)\Eh+\fb(\zb)\Et+\frac{1}{\sqrt{2}} E_- \label{Mminusdiffeq2}.
\eea
This statement of the solution makes sense in the context of $M_+^{-1} M_-$ which satisfy the decomposition $M_+^{-1} M_-=N_- N_0 N_+^{-1}$.  As before, we will assume that this decomposition is possible for the exponentiated ``group.''

To begin with, we will consider the differential relations
\be
\frac{1}{2i}\pa\Lambda_2=\frac{1}{2i}\tanh^2(\phi_1)\pa\Lambda_1, \quad \frac{1}{2i}\pab\Lambda_2=-\frac{1}{2i}\tanh^2(\phi_1)\pab\Lambda_1
\ee
We consider first the holomorphic derivative constraint, and write this out in terms of the $\la i j \ra$, and find
\be
\frac{\pa \la 33 \ra \la 11\ra - \pa \la 11 \ra \la 33 \ra -\pa\la 31\ra \la 13 \ra+ \pa \la 13 \ra \la 31 \ra  }{\la 11 \ra \la 33 \ra}=0 \label{laholo1}
\ee
Let us examine the terms that appear above:
\bea
\pa\la i j\ra &=& \la i \mid \pa M_+^{-1} M_- \mid j \ra \nn \\
&=& \la i \mid \big(\pa h_1(z)H_1+\pa h_2(z)H_2+\pa g_{\D}(z)\D+\pa g_{\kh}\kh \nn \\
&& \qquad + f_{\wedge}(z) \Eh + f_{\sim} \Et+\frac{1}{\sqrt{2}} E_+\big)M_+^{-1} M_-^{-1}\mid j \ra. \label{iplusexpand1}
\eea
Now, we know exactly how $H_i, \D, \kh, \Eh, \Et$ work on all of the components $\la i\mid$ for any $i$, and so one can simply read off these component by component.  For example, plugging (\ref{iplusexpand1}) into (\ref{laholo1}), and paying attention only to the $H_1$ insertion, we find
\be
\frac{\pa h_1}{2}\left(
\frac{\la 13 \ra \la 11\ra - \la 31 \ra \la 33 \ra -\la 11\ra \la 13 \ra+  \la 33 \ra \la 31 \ra  }{\la 11 \ra \la 33 \ra} \right)=0.
\ee
Similarly, one can track the $H_2, \D, \kh$ and $\Et$ insertions and see that they all vanish.  However, the insertion of $\Eh$ gives
\be
if_{\wedge}\left(\frac{- \la 33 \ra \la 11\ra - \pa \la 11 \ra \la 33 \ra +\pa\la 31\ra \la 13 \ra+ \pa \la 13 \ra \la 31 \ra  }{\la 11 \ra \la 33 \ra}\right)\neq 0.
\ee
Hence, we see that to satisfy the constraint equation, we must set $f_{\wedge}=0$.  Finally, this leaves the insertions of $E_{+}$ to consider.  We give new notation, and define the following
\bea
&& \la i_+ j\ra \equiv \la i \mi E_+ M_+^{-1} M_- \mi j\ra, \quad \la i j_-\ra= \la i \mi M_+^{-1} M_- E_- \mi j\ra \nn \\
&& \qquad \qquad \qquad \qquad\la i_+ j_-\ra= \la i \mi E_+ M_+^{-1} M_- E_- \mi j\ra.
\eea
We will find this notation convenient later on.  However, now we are left evaluating
\be
\frac{1}{\sqrt{2}}\frac{\la 3_+3 \ra \la 11\ra - \la 1_+ 1 \ra \la 33 \ra -\la 3_+1\ra \la 13 \ra+  \la 1_+3 \ra \la 31 \ra  }{\la 11 \ra \la 33 \ra}=0.
\ee
To evaluate the above, we note that we may replace $M_+^{-1} M_-$ by $N_- \cE N_+^{-1}$.  Then, we see that
\be
\la i_+ j\ra=\la i \mid E_+ N_- \cE N_+ \mi j \ra=\la i \mid E_+ N_- \cE \mi j \ra
\ee
where $N_+\mid j\ra=\mid j\ra$ because $\mi j \ra$ is highest grade.  We take advantage of the fact that $N_-$ is an exponentiated generator of ${\mc N}_-$ and so we see that $N_-=1+n_{-}+\cdots$ where the $\cdots$ denote terms that are grade $-2$ or less.  Terms that have grade $-2$ or less will annihilate $\la i \mi E_+$ because this state is grade $-1$, and the highest grade is grade 0. Therefore, it is only the grade $-1$ part of the generator that matters, which we have denoted $n_-$.  Therefore, we see that
\be
\la i_+ j\ra=\la i \mid E_+ N_- \cE N_+ \mi j \ra=\la i \mid E_+ N_- \cE \mi j \ra=\la i \mid E_+ (1+n_-) \cE \mi j \ra.
\ee
Next, we see that the grade of the left hand and right hand vectors only match when we take the $n_-$ term, and ignore the $1$ in the expansion (inner products between vectors of different grades is zero).  Therefore, we finally see that
\be
\la i_+ j\ra=\la i \mid E_+ n_- \cE \mi j \ra=\la i_+ j\ra=\la i \mid [E_+, n_-] \cE \mi j \ra
\ee
where we replace with the commutator because $\la j \mi n_-=0$.  We take the general form of the grade $-1$ operator in the subalgebra $\hat{s}_4$
\bea
n_-&=&n_{-1}E_{[1,1,0]}^{-1}+n_{-2}E_{[-1,1,0]}^{-1}\nn \\
&& \qquad \qquad +n_{-3}(E_{[0,-1,1]}^{-1}+E_{[0,-1,-1]}^{-1})
+n_{-4}(E_{[0,1,1]}^{-1}-E_{[0,1,-1]}^{-1}).
\eea
The commutator is easy to evaluate, and we find
\be
[E_+, n_-]=U n_{-1}(-H_1-H_2+\kh)+U n_{-2}(H_1-H_2+\kh)+n_{-3}(2H_2+2\kh)-U n_{-4}\Et.
\ee
We can evaluate how this acts as on $\la i \mi$ for all of the $i=1,2,3,4$.  We see that
\bea
&& \la 1 \mi E_+ n_-=A_+ \la 1 \mi + C_{+}\la 3 \mi \quad \la 3 \mi E_+ n_-=A_+\la 3 \mi + B_+ \la 1 \mi \nn \\
&& \la 2 \mi E_+ n_-=D_+ \la 2 \mi + C_{+}\la 4 \mi \quad \la 3 \mi E_+ n_-=D_+\la 3 \mi + B_+ \la 1 \mi
\eea
with
\bea
A_+=\frac{U n_{-1}}{2}+\frac{U n_{-2}}{2}+\frac{6 n_{-3}}{2}, \quad B_+=-\frac{U n_{-1}}{2}+\frac{U n_{-2}}{2}-\frac{2i n_{-4}}{2} \nn \\
D_+=\frac{3U n_{-1}}{2}+\frac{3U n_{-2}}{2}+\frac{2 n_{-3}}{2}, \quad C_+=-\frac{U n_{-1}}{2}+\frac{U n_{-2}}{2}+\frac{2i n_{-4}}{2}
\eea
Similarly, one can show that for
\be
n_+=n_{+1}E_{[-1,-1,0]}^{1}+n_{+2}E_{[1,-1,0]}+n_{+3}(E_{[0,1,-1]}^1+E_{[0,1,1]}^1)+n_{+4}(E_{[0,-1,-1]}^1-E_{[0,-1,1]}^1)
\ee
we have
\bea
&& n_+ E_- \mi 1 \ra =A_- \mi 1 \ra + B_{-}\mi 3 \ra \quad n_+ E_-\mi 3 \ra =A_-\mi 3 \ra + C_- \mi 1 \ra \nn \\
&& n_+ E_- \mi 2 \ra =D_- \mi 2 \ra + B_{-}\mi 4 \ra \quad n_+ E_-\mi 4 \ra =D_-\mi 4 \ra + C_- \mi 2 \ra
\eea
with
\bea
A_-=\frac{\Ub n_{+1}}{2}+\frac{\Ub n_{+2}}{2}+\frac{6 n_{+3}}{2}, \quad B_-=-\frac{\Ub n_{+1}}{2}+\frac{\Ub n_{+2}}{2}-\frac{2i n_{+4}}{2} \nn \\
D_-=\frac{3\Ub n_{+1}}{2}+\frac{3\Ub n_{+2}}{2}+\frac{2 n_{+3}}{2}, \quad C_-=-\frac{\Ub n_{+1}}{2}+\frac{\Ub n_{+2}}{2}+\frac{2i n_{+4}}{2}.
\eea
Finally, because the $\la i \mi M_+^{-1} M_{-} \mi j \ra=\la i \mi \cE \mi j \ra$ we see that we may again read off coefficients.  We find
\bea
&& \frac{1}{\sqrt{2}}\frac{\la 3_+3 \ra \la 11\ra - \la 1_+ 1 \ra \la 33 \ra -\la 3_+1\ra \la 13 \ra+  \la 1_+3 \ra \la 31 \ra  }{\la 11 \ra \la 33 \ra}\nn \\
&&\qquad =\frac{1}{\sqrt{2}\la 11 \ra \la 33 \ra} \Bigg(A_+ \left(\la33\ra\la11\ra-\la11\ra\la33\ra-\la31\ra\la13\ra+\la13\ra\la31\ra \right) \nn \\
&&\qquad \quad +B_+(\la13\ra\la11\ra-\la11\ra\la13\ra)+ C_+(-\la31\ra\la33\ra+\la33\ra\la31\ra)\Bigg)=0.
\eea
Hence we see that $\pa \Lambda_2=\tanh^2(\phi_1) \pa \Lambda_2$ is true iff $f_{\wedge}=0$.

One may do likewise for the antiholomorphic constraint $\pab \Lambda_2=-\tanh^2{\phi_1}\pab\Lambda_1$ and write
\be
\frac{\pab \la 33 \ra \la 11\ra - \pab \la 11 \ra \la 33 \ra +\pab\la 31\ra \la 13 \ra- \pab \la 13 \ra \la 31 \ra  }{\la 11 \ra \la 33 \ra}=0 \label{laantiholo1}.
\ee
Plugging in instead the $M_-$ equation of motion, we find again that the terms coming from $H_i, \D, \kh, \Et$ insertions vanish identically.  However, again the $\Eh$ insertion, multiplying $\bar{f}_{\wedge}$, does not vanish and so we must take $\bar{f}_{\wedge}=0$.  We may then consider the $E_-$ insertions, and then expand in the functions $A_-, B_-, C_-, D_-$ and again see that these coefficients vanish.  Hence, we see that the constraints
\bea
&& \frac{1}{2i}\pa\Lambda_2=\frac{1}{2i}\tanh^2(\phi_1)\pa\Lambda_1 \leftrightarrow f_{\wedge}=0, \nn \\
&& \frac{1}{2i}\pab\Lambda_2=-\frac{1}{2i}\tanh^2(\phi_1)\pab\Lambda_1 \leftrightarrow \bar{f}_{\wedge}=0.
\eea
Since these constraints in fact give the $\Lambda_1$ equation of motion, we see that the $\Lambda_1$ equation of motion is satisfied.  In the process we have also removed the unphysical degree of freedom from the earlier function count, and reduced to five holomorphic and five antiholomorphic functions: the correct count for five fields.

The other equations of motion follow from similar considerations.  To illustrate this, we will consider the simplest of the equations: the equation for $\phi_2$.  This will have all relevant features for evaluating the other equations of motion as well.  The $\nu$ equation of motion and, particularly, the $\phi_1$ equation of motion require similar reasoning, just with more tedious algebra.

To check the $\phi_2$ equation of motion, we start with
\bea
2\phi_2&=&\ln\left(\frac{\la 24 \ra}{\la 13\ra}\right) \nn \\
2\pab \phi_2&=& \frac{\pab \la 24\ra}{\la 24\ra}- \frac{\pab \la 13 \ra}{\la 13 \ra}.
\eea
We expand as we did before, using various insertions, and find
\bea
2\pab \phi_2&=& -\frac{\pab \hb_1}{2}\left(\frac{\la 22 \ra }{\la 24\ra}-\frac{\la 11 \ra}{\la 13 \ra}\right)-\frac{\pab \hb_2}{2}\left(-\frac{\la 24 \ra }{\la 24\ra}-\frac{\la 13 \ra}{\la 13 \ra}\right) \nn \\
&&+i\fb \left(-\frac{\la 22 \ra }{\la 24\ra}+\frac{\la 11 \ra}{\la 13 \ra}\right) -\frac{1}{\sqrt{2}}\left(\frac{ \la 24_-\ra}{\la 24\ra}- \frac{\la 13_- \ra}{\la 13 \ra}\right) \nn \\
&=& \pab \hb_2 -\frac{1}{\sqrt{2}}\left(\frac{ \la 24_-\ra}{\la 24\ra}- \frac{\la 13_- \ra}{\la 13 \ra}\right)
\eea
where the terms that have vanished do so because of the redundancies (\ref{redundant1}),(\ref{redundant2}).  Note that we will not expand out the terms $\la i j_-\ra$ simply because the functions $A_\pm, B_\pm, C_\pm, D_\pm$ do not satisfy any particularly nice holomorphicity conditions, while the $E_\pm$ do.  Taking the $z$ derivative, we find
\bea
2\pa\pab \phi_2 &=& \frac{1}{\sqrt{2}}\left(\frac{\pa \la 24_-\ra\la24\ra-\pa\la24\ra\la24_-\ra}{\la 24\ra^2}-\frac{\pa \la 13_-\ra\la13\ra-\pa\la13\ra\la13_-\ra}{\la 24\ra^2}\right).
\eea
This is likewise easy to expand using the insertions of various types
\bea
2\pa\pab \phi_2= -\frac{1}{\sqrt{2}}\Bigg[&&\!\frac{\pa h_1}{2}\left(\frac{\la 44_-\ra\la24\ra-\la44\ra\la24_-\ra}{\la 24\ra^2}-\frac{\la 33_-\ra\la13\ra-\la33\ra\la13_-\ra}{\la 24\ra^2}\right) \nn \\
&&\!+\frac{\pa h_2}{2} \left(-\frac{\la 24_-\ra\la24\ra-\la24\ra\la24_-\ra}{\la 24\ra^2}-\frac{\la 13_-\ra\la13\ra-\la13\ra\la13_-\ra}{\la 24\ra^2}\right) \nn \\
&&\!+if \left(-\frac{\la 44_-\ra\la24\ra-\la44\ra\la24_-\ra}{\la 24\ra^2}+\frac{\la 33_-\ra\la13\ra-\la33\ra\la13_-\ra}{\la 24\ra^2}\right) \\
&&\!+\frac{1}{\sqrt{2}}\left(\frac{\la2_+4_-\ra\la24\ra-\la2_+4\ra\la24_-\ra}{\la24\ra} -\frac{\la1_+3_-\ra\la13\ra-\la1_+3\ra\la13_-\ra}{\la13\ra}\right)\Bigg] \nn
\eea
We note that the coefficient of $\pa h_1/2$ and $if$ are the same up to a minus sign, and we look at these terms to find
\bea
&& \frac{\la 44_-\ra\la24\ra-\la44\ra\la24_-\ra}{\la 24\ra^2}-\frac{\la 33_-\ra\la13\ra-\la33\ra\la13_-\ra}{\la 24\ra^2} \nn \\
&&= D_-\frac{\la 44\ra \la24-\la44\ra\la24\ra}{\la24\ra^2}+C_-\frac{\la42\ra\la24\ra-\la44\ra\la22\ra}{\la24\ra^2} \nn \\
&& -A_-\frac{\la33\ra\la13\ra-\la33\ra\la13\ra}{\la13\ra^2}-C_-\frac{\la31\ra\la13\ra-\la33\ra\la13\ra}{\la13\ra^2} \nn \\
&&= C_-\left(\frac{\la42\ra}{\la24\ra}-\frac{\la31\ra}{\la13\ra}-\frac{\la 44\ra}{\la22\ra}\left(\frac{\la 22\ra}{\la 24\ra}\right)^2+\frac{\la 33\ra}{\la11\ra}\left(\frac{\la11\ra}{\la13\ra}\right)^2\right)=0
\eea
where in the last line we have made use of the redundancies (\ref{redundant1}),(\ref{redundant2}).  So we find that
\be
-4\pa\pab \phi_2=\left(\frac{\la2_+4_-\ra\la24\ra-\la2_+4\ra\la24_-\ra}{\la24\ra} -\frac{\la1_+3_-\ra\la13\ra-\la1_+3\ra\la13_-\ra}{\la13\ra}\right).
\ee
There is one final game to play to evaluate the above expression.  We need to evaluate terms of the sort $\la i_+ j_-\ra$, which we have not dealt with before.  However, we again use the relation (\ref{cedef}) and find
\be
\la i_+ j_-\ra=\la i \mid E_+ M_+^{-1} M_- E_-\mid j \ra= \la i \mi E_+ N_- \cE N_+^{-1}E_-\mi j\ra
\ee
We may again expand $N_-=(1+n_-)$ and $N_+^{-1}=1+n_+$, and evaluate terms.  In fact, it is easy to see that we must either insert the $1$s simultaneously or the $n_\pm$ simultaneously: other terms will have mismatched grades and so the inner product is 0.  We therefore invent one more layer of notation, and call
\be
\la i \mi E_+ \cE E_-\mi j\ra\equiv \la i_+ j_-\ra_\cE.
\ee
For the other terms, where the $n_\pm$ are inserted, we may expand the $n_+ E_- \mid j \ra$ and $\la i \mid n_- E_+$ using $A_\pm, B_\pm, C_\pm, D_\pm$ as before.  Hence, we find that
\bea
-4\pa \pab \phi_2&&\! = \frac{\la 2_+ 4_-\ra_\cE}{\la 24\ra}- \frac{\la 1_+ 3_-\ra_\cE}{\la 13\ra}\nn \\
&&+\Bigg[ D_+D_-\frac{\la 24 \ra \la24\ra -\la24\ra\la24\ra}{\la24\ra^2}+ D_+C_-\frac{\la22\ra\la24\ra-\la24\ra\la22\ra}{\la24\ra^2}\nn \\
&&C_+D_-\frac{\la44\ra\la24\ra-\la44\ra\la24\ra}{\la24\ra^2}+C_+C_-\frac{\la42\ra\la24\ra-\la44\ra\la22\ra}{\la24\ra^2} \nn \\
&&-A_+A_-\frac{\la 13 \ra \la13\ra -\la13\ra\la13\ra}{\la13\ra^2} -A_+C_-\frac{\la11\ra\la13\ra-\la13\ra\la11\ra}{\la13\ra^2}\nn \\
&&-C_+A_-\frac{\la33\ra\la13\ra-\la33\ra\la13\ra}{\la13\ra^2}-C_+C_-\frac{\la31\ra\la13\ra-\la33\ra\la11\ra}{\la13\ra^2}
\Bigg] \nn \\
=&&\frac{\la 2_+ 4_-\ra_\cE}{\la 24\ra}- \frac{\la 1_+ 3_-\ra_\cE}{\la 13\ra} \label{phi2eps}
\eea
where we cancel almost all of the terms using the redundancies (\ref{redundant1}), (\ref{redundant2}).

We are now left to contend with expressions of the form $\la i_+ j_-\ra_\cE$.  First, we expand this out
\be
\la i \mi E_+ \cE E_- \mi j \ra = \la i \mi E_+ e^{-\Lambda_+}e^{-2\phi}e^{\Lambda_-} E_- \mi j \ra.
\ee
On the right hand side of the above expression, we recall that $[E_{\pm},\Lambda_{\pm}]=0$ for any assignment of the subscripts.  Thus, the $\Eh$ eigenvalues have been unaffected.  We will refer to the eigenvalues as $\Eh\mi i \ra= \lambda_i \mi i \ra$.  After removing the $\Lambda_\pm$ in this way, we see that the remaining vectors are also $\D$ eigenvectors, with eigenvalue $-1$ (coming from $\D$ commuting grade $-1$ operators on the right, or grade $+1$ on the left), and as always we may also evaluate the $\kh$ eigenvalue which is never affected.  Hence, we see that
\be
\la i_+ j_-\ra_\cE=e^{-\lambda_i\frac{\Lambda_1+\Lambda_2}{2}}e^{\lambda_j\frac{\Lambda_1-\Lambda_2}{2}}e^{2\eta}e^{-2\nu}\la i\mi E_+ e^{-2\phi_0}E_-\mi j\ra
\ee
where $\phi_0=\phi_1 H_1^0+\phi_2 H_2^0$.  We rewrite this final expression as
\bea
\la i_+ j_-\ra_\cE=e^{-\lambda_i\frac{\Lambda_1+\Lambda_2}{2}}e^{\lambda_j\frac{\Lambda_1-\Lambda_2}{2}}e^{2\eta}e^{-2\nu}\la i\mi [E_+, e^{-2\phi_0}E_-e^{2\phi_0}]e^{2\phi_0}\mi j\ra
\eea
which we may do because $E_+$ annihilates any highest grade state; e.g. $E_+ e^{2\phi_0}\mi{i}\ra=0$.  The commutator $[E_+, e^{-2\phi_0}E_-e^{2\phi_0}]$ will have only grade 0 operators in them, including some occurrences of $\kh$ which is set to $1$ for this entire representation.  Hence, the above inner product may in fact be understood in terms of the base Lie algebra and base Lie group.  This expression is
\bea
&& \! \! \la i\mi [E_+, e^{-2\phi_0}E_-e^{2\phi_0}]e^{2\phi_0}\mi j\ra \nn \\
&&=\la i \mi
\Big(\Ub U e^{-2\phi_1-2\phi_2}(-H_1-H_2+1) \nn \\
&& \kern4em+\Ub U e^{2\phi_1-2\phi_2}(H_1-H_2+1)
+e^{2\phi_2}(2H_2+2)\Big)e^{2\phi_0}\mi j\ra.
\eea
Using this expression, the explicit representation in the appendix, and the definition of the $\mid i\ra$ vectors, and along with the values of $\la i j\ra$ we find that (\ref{phi2eps}) becomes
\be
-4\pa \pab \phi_2=-e^{2\eta}\left(-2U\Ub e^{-2\phi_2}\cosh(2\phi_1)+2e^{2\phi_2}\right),
\ee
and so the equation of motion is satisfied.  Similarly, but with much more algebra, one can show that
\bea
8\pa \pab \nu &=& \frac{\la2_+2_-\ra_\cE\la33\ra+\la3_+3_-\ra_\cE\la22\ra-\la1_+3_-\ra_\cE\la42\ra-\la4_+2_ -\ra_\cE\la13\ra}{\la22\ra\la33\ra-\la13\ra\la42\ra} \nn \\
&=&e^{2\eta}\left(4U\Ub\cosh(2\phi_1)e^{-2\phi_2}+4e^{2\phi_2}\right)
\eea
and so this equation of motion is also satisfied.  Finally, with a copious amount of algebra, one can show
\bea
&& 2\pa \pab \phi_1 -\frac{\pa\Lambda_1 \pab\Lambda_2\sinh(2\phi_1)}{\cosh^4(\phi_1)}\nn \\&& =-\frac{1}{2\left(\sqrt{\frac{\la 11\ra\la33\ra}{\la13\ra\la31\ra}}-\sqrt{\frac{\la13\ra\la31\ra}{\la 11\ra\la33\ra}}\right)}\left[\frac{\la1_+3_-\ra_\cE}{\la13\ra}+\frac{\la3_+1_-\ra_\cE}{\la31\ra} -\frac{\la1_+1_-\ra_\cE}{\la11\ra}-\frac{\la3_+3_-\ra_\cE}{\la33\ra}\right] \nn \\
&&=U\Ub e^{2\eta}e^{-2\phi_2}\sinh(2\phi_1)
\eea
and so this equation of motion is satisfied as well.


\section{Discussion and conclusions}
\label{discussion}

We have shown above that the Pohlmeyer reduced sigma model describing minimal area surfaces in AdS$_5$ may be conformally extended to the model presented in section 2.  This model admits a Lax pair valued in the affine Lie algebra $\widehat{su(4)}$, or more precisely, the ${\mathbb Z}_4$ symmetric subalgebra $\hat{s}_4$.  It would be interesting to know the significance of this subalgebra, and characterize it more completely.

One reason to do this is because of the analysis of section 3.  In this section 3, we have relied on a decomposition, which we have simply assumed.  The decomposition is clear at the level of the algebra, however, whether it descends to $\exp(\hat{s}_4)$ seems to us to be non-trivial.  Assuming these decompositions are general, we have been able to show that the general form of the solution is that of (\ref{cedef}), and further that this form of the solution solves the equations of motion.  We do not address whether this form for the solution is sensible, however, it again follows from assuming that the decomposition described in section 3 is generically possible.  Hence, under the assumption that the structure of the algebra as $\hat{s}_4={\mathcal N}_- \oplus {\mathcal N}_0 \oplus {\mathcal N}_+$ descends to the exponentiated elements as $g=N_- N_0 N_+$ we see that the form of the solution (\ref{cedef}) is equivalent to solving the equations of motion: i.e. these are in principle general solutions.

Given that these are in fact general solutions, one should be able to use the above formalism to develop an algorithm for determining them to very good accuracy.  The path ordered exponential is naturally defined in terms of discretizing the ``steps'' of the integral, and so may be thought of as an algorithm which converges when the number of steps is taken to $\infty$.  Also, because the solutions are written in terms of the highest grade states only, one could hope that the ``depth'' into the representation that the exponentiated terms reach is exponentially suppressed.  If this is true, it implies that if one could get reasonable approximations to the solutions by taking a sufficiently large piece of the highest weight representation, along with the path ordered exponential approximated by a large, but finite number of steps.  Thus, one could in principle ``experimentally'' test the validity of the above solutions, by testing whether they approximately satisfy the equations of motion.

One problem with these formal solutions, as always, is understanding the map from boundary conditions to the data that is naturally specified in the solution: for us, the holomorphic and antiholomorphic functions defining $M_{+}^{-1} M_{-}$.  Never the less, it would be interesting to see if there were classes of known special functions that could possibly solve the equations of motion, as recently discussed in \cite{Ishizeki:2011bf}, using results from \cite{Babich:1992mc}.  This may be relatively difficult, but there is some guidance given by considering the AdS$_3$ case which could lead to the equations discussed in \cite{Ishizeki:2011bf}.  Further, this may be possible due to the connection between theta functions and affine Lie algebras (see chapter 13 of \cite{Kac:1990gs}) on which the solutions studied in \cite{Babich:1992mc,Ishizeki:2011bf} are based.

One may wonder what other types of systems admit the analysis that we have done in section 3.  There is one candidate worth exploring: theories based on $\widehat{su(n)}$.  The basic feature of the $\widehat{su(4)}$ model is that it exists in some ${\mathbb Z}_4$-invariant subalgebra.  This is natural because the base level of the ladder operators in $E_+$ are all congruent modulo 4.  In the $\widehat{su(n)}$ case, the generalization of $E_+$ is constructed using operators that have base level congruent modulo $n$ and so some ${\mathbb Z}_n$ reduction of $\widehat{su(n)}$ could in principle be made.  One can be further guided by the $su(4)$ case discussed here by noting that $\Eh\propto E_+^2$ when restricting to the base lie algebra and considering the the defining ${\bf 4}$ representation.  Hence, in the $su(n)$ case it seems natural to consider the other matrix structures of a similar form, i.e. $E_+^k$, and give these similar looking ``non commuting'' kinetic terms in the Lax pair.

Finally, we also comment on some results from appendix \ref{dimreduct}.  Here we write a system that is a dimensional reduction of the system described in the text which retains the integrability.  This reduced system should be able to be solved by quadratures, and possibly give new types of minimal surfaces that truly live in all of AdS$_5$, rather than just an AdS$_3$ subspace. We will leave this question and those mentioned above to future work.

\section*{Acknowledgements}
I am grateful to Peng Gao for conversations early on in this work.
The author is funded under a grant from NSERC Canada, and receives additional support from IPP Canada.

\appendix

\section{Representations for Lie algebras $su(4)=so(6)$.}
\label{repsforsu4}

We display here a possible representation of the Lie algebra ${\rm A}_3={\rm D}_3$.  For this we give first the Cartan generators
\bea
H_1=\begin{pmatrix}1 & 0 & 0 & 0 & 0 & 0 \\ 0 & -1 & 0 & 0 & 0 & 0 \\ 0 & 0 & 0 & 0 & 0 & 0\\
0 & 0 & 0 & 0 & 0 & 0\\0 & 0 & 0 & 0 & 0 & 0\\0 & 0 & 0 & 0 & 0 & 0\\ \end{pmatrix},&& \quad
H_2=\begin{pmatrix}0 & 0 & 0 & 0 & 0 & 0 \\ 0 & 0 & 0 & 0 & 0 & 0 \\ 0 & 0 & 1 & 0 & 0 & 0\\
0 & 0 & 0 & -1 & 0 & 0\\0 & 0 & 0 & 0 & 0 & 0\\0 & 0 & 0 & 0 & 0 & 0\\ \end{pmatrix}, \quad
 H_3=\begin{pmatrix}0 & 0 & 0 & 0 & 0 & 0 \\ 0 & 0 & 0 & 0 & 0 & 0 \\ 0 & 0 & 0 & 0 & 0 & 0\\
0 & 0 & 0 & 0 & 0 & 0\\0 & 0 & 0 & 0 & 1 & 0\\0 & 0 & 0 & 0 & 0 & -1\\ \end{pmatrix}
\eea
and then the generators associated with the positive simple roots
\bea
E_{[1,-1,0]}=\begin{pmatrix}0 & 0 & 1 & 0 & 0 & 0 \\ 0 & 0 & 0 & 0 & 0 & 0 \\ 0 & 0 & 0 & 0 & 0 & 0\\
0 & -1 & 0 & 0 & 0 & 0\\0 & 0 & 0 & 0 & 0 & 0\\0 & 0 & 0 & 0 & 0 & 0\\ \end{pmatrix},&& \quad
E_{[0,1,-1]}=\begin{pmatrix}0 & 0 & 0 & 0 & 0 & 0 \\ 0 & 0 & 0 & 0 & 0 & 0 \\ 0 & 0 & 0 & 0 & 1 & 0\\
0 & 0 & 0 & 0 & 0 & 0\\0 & 0 & 0 & 0 & 0 & 0\\0 & 0 & 0 & -1 & 0 & 0\\ \end{pmatrix} \nn \\
&& \kern-5em E_{[0,1,1]}=\begin{pmatrix}0 & 0 & 0 & 0 & 0 & 0 \\ 0 & 0 & 0 & 0 & 0 & 0 \\ 0 & 0 & 0 & 0 & 0 & 1\\
0 & 0 & 0 & 0 & 0 & 0\\0 & 0 & 0 & -1 & 0 & 0\\0 & 0 & 0 & 0 & 0 & 0\\ \end{pmatrix}
\eea
We further define
\bea
&& E_{[1,0,-1]}\equiv \left[E_{[1,-1,0]},E_{[0,1,-1]}\right],\nn \\ \quad
&&E_{[1,0,1]}\equiv -\left[E_{[1,-1,0]},E_{[0,1,1]}\right], \label{higherviacom}\\
&&\quad E_{[1,1,0]}\equiv \left[E_{[1,0,-1]},E_{[0,1,1]}\right]=\left[\left[E_{[1,-1,0]},E_{[0,1,-1]}\right],E_{[0,1,1]}\right]. \nn
\eea
Of course any signs could be used above.  The above has defined all positive roots, and to define the negative roots we take $E_{[-i,-j,-k]}=\left(E_{[i,j,k]}\right)^T$, where $^T$ denotes the transpose.

There also exists a $4\times 4$ representation of the Lie algebra $A_3=D_3$.
\bea
&& H_1=\begin{pmatrix} \frac12 & 0 & 0 & 0 \\ 0 & \frac12 & 0 & 0 \\ 0 & 0 & -\frac12 & 0 \\ 0 & 0 & 0 & -\frac12\end{pmatrix},  \qquad H_2= \begin{pmatrix} \frac12 & 0 & 0 & 0 \\ 0 & -\frac12 & 0 & 0 \\ 0 & 0 & \frac12 & 0 \\ 0 & 0 & 0 & -\frac12\end{pmatrix}, \qquad H_3=\begin{pmatrix} -\frac12 & 0 & 0 & 0 \\ 0 & \frac12 & 0 & 0 \\ 0 & 0 & \frac12 & 0 \\ 0 & 0 & 0 & -\frac12\end{pmatrix}\\
&& E_{[1,-1,0]}=\begin{pmatrix} 0 & 0 & 0 & 0 \\ 0 & 0 & 1 & 0 \\ 0 & 0 & 0 & 0 \\ 0 & 0 & 0 & 0 \end{pmatrix}, \qquad
E_{[0,1,-1]}= \begin{pmatrix} 0 & 1 & 0 & 0 \\ 0 & 0 & 0 & 0 \\ 0 & 0 & 0 & 0 \\ 0 & 0 & 0 & 0 \end{pmatrix}, \qquad E_{[0,1,1]}= \begin{pmatrix} 0 & 0 & 0 & 0 \\ 0 & 0 & 0 & 0 \\ 0 & 0 & 0 & 1 \\ 0 & 0 & 0 & 0 \end{pmatrix}
\eea
along with the same definitions (\ref{higherviacom}) for the positive roots, and $E_{[-i,-j,-k]}=\left(E_{[i,j,k]}\right)^T$ for the negative roots.

\section{How obtain the $\hat{s}_4$ model as a reduction of the general $\widehat{su(4)}$ model.}

\label{reductionviasymmetry}

Let us use the transformations of section \ref{confextend} to take the general model and reduce it to the $\hat{s}_4$ form.  First, to define the general model, we must first specify the pole structure in the spectral parameter.  We take that the Lax connection $A$ has a simple pole at $\lambda=\infty$ and $\Ab$ has a simple pole at $\lambda=0$.  Therefore, the Lax connection has the form
\be
A(\lambda)=A_0+\lambda A_1, \quad \Ab(\lambda)=\Ab_0+\frac{1}{\lambda} \Ab_{-1}. \label{asplit}
\ee
Further, we take that the connection is valued in the affine Lie algebra $\widehat{su(4)}$.

To reduce the theory, let us recall that the Lax pair equations of motion are that the field strength associated with $A(\lambda),\Ab(\lambda)$ is zero.  Hence if we have a solution to the zero curvature equations $A(\lambda),\Ab(\lambda)$, then $A'(\lambda)=M^{-1}A(\lambda) M,\Ab'(\lambda)=M^{-1}\Ab(\lambda) M$ must also be a solution to the equations of motion for any constant invertible matrix $M$ (we also require that $M$ does not depend on $\lambda$).  However, requiring that $A'(\lambda)=A(\lambda),\Ab'(\lambda)=\Ab(\lambda)$ can be too stringent, as the zero curvature condition must hold as functions of $\lambda$, and so any linear function of $\lambda$ could appear in the argument of $A'$ and $\Ab'$.
Therefore, to reduce a theory, one takes the subset of Lax connections $A(\lambda),\Ab(\lambda)$ that obey some symmetry $G$,
\bea
{\mathcal R}^{-1} A(r(\lambda)) {\mathcal R}=A(\lambda), \quad {\mathcal R}^{-1} \Ab(r(\lambda)) {\mathcal R}=\Ab(\lambda) \label{genreduction}
\eea
where ${\mathcal R}$ is an appropriate representation of the group $G$ and $r$ is a one (complex) dimensional representation of $G$.  Using the above, we may first make the restriction
\be
C^{-1}_U(\theta) A\left(e^{-i\theta}\lambda\right) C_U(\theta)=A\left(\lambda\right). \label{u1restriction2}
\ee
and so the first group that we require symmetry under is the $U(1)$ mentioned in the text.

The restriction (\ref{u1restriction2}) says that the power of $\lambda$ multiplying the generator determines the the superscript $^n$.  The only elements that may appear are $M^n$ with $-1\leq n \leq 1$.  This simply says that $A_0$ and $\Ab_0$ are grade $0$ operators while $A_{1}$ is grade $1$ and $\Ab_{-1}$ is grade $-1$.  Now, we further require that $P(A)=A$ and $P(\Ab)=\Ab$.  This restricts the tensor structure to be those already listed in the main text, i.e.
\bea
&& A_0\sim \left[H_1^0, H_2^0, \D , \kh, E_{[1,0,1]}^{0}+ E_{[1,0,-1]}^{0}, E_{[-1,0,1]}^0+ E_{[-1,0,-1]}^0 \right] \\
&& A_1\sim \left[ E_{[1,-1,0]}^{1}, E_{[0,1,-1]}^{1}+ E_{[0,1,1]}^{1}, E_{[-1,-1,0]}^{1},E_{[0,-1,-1]}^1-E_{[0,-1,1]}^1\right] \\
&& \Ab_0\sim \left[H_1^0, H_2^0, \D , \kh, E_{[1,0,1]}^{0}+ E_{[1,0,-1]}^{0}, E_{[-1,0,1]}^0+ E_{[-1,0,-1]}^0 \right] \\
&& \Ab_{-1} \sim \left[ E_{[-1,1,0]}^{-1}, E_{[0,-1,1]}^{-1}+ E_{[0,-1,-1]}^{-1}, E_{[1,1,0]}^{-1},E_{[0,1,1]}^{-1}-E_{[0,1,-1]}^{-1}\right]
\eea
(note that this has removed $H_3^0$).  So, the generator form has almost been reduced to that of our Lax connection.

Let us write out the Lax connection in terms of these general structure
\bea
A&&=-\pa\beta_{01} H_1^0-\Phi_2 H_2^0-\pa \eta \D -\pa \nu \kh\nn\\
&&-\frac12\left(\pa\beta_{02}-\pa\beta_{12}\right)\left(E_{[1,0,-1]}^{0}+E_{[1,0,1]}^{0}\right)
  -\frac12\left(\pa\beta_{02}+\pa\beta_{12}\right)\left(E_{[-1,0,-1]}^{0}+E_{[-1,0,1]}^{0}\right) \nn \\
&&+\frac{\lambda}{\sqrt{2}}e^{-\phib_2+\etab}\left(u_0+u_1\right) E_{[-1,-1,0]}^1 + \frac{\lambda}{\sqrt{2}}e^{-\phib_2+\etab}\left(u_0-u_1\right) E_{[1,-1,0]}^1 \nn \\
&&+\frac{\lambda}{\sqrt{2}}e^{\phib_2+\etab}\left(E_{[0,1,1]}^1+E_{[0,1,-1]}^1\right)-\frac{\lambda}{\sqrt{2}}u_2 e^{-\phib_2+\etab}\left(E_{[0,-1,1]}^1-E_{[0,-1,-1]}^1\right) \\
\Ab&&=-\pab \betab_{01} H_1^0+\Phib_2 H_2^0 +\pab \etab \D +\pab \nub \kh \nn\\
&&-\frac12\left(\pab\betab_{02}+\pab\betab_{12}\right)\left(E_{[-1,0,-1]}^{0}+E_{[-1,0,1]}^{0}\right)
  -\frac12\left(\pab\betab_{02}-\pab\betab_{12}\right)\left(E_{[1,0,1]}^{0}+E_{[1,0,-1]}^{0}\right) \nn \\
&&+\frac{1}{\lambda\sqrt{2}}e^{-\phi_2+\eta}\left(\ub_0-\ub_1\right) E_{[1,1,0]}^{-1} + \frac{1}{\lambda\sqrt{2}}e^{-\phi_2+\eta}\left(\ub_0+\ub_1\right) E_{[-1,1,0]}^{-1} \nn \\
&&+\frac{1}{\lambda\sqrt{2}}e^{\phi_2+\eta}\left(E_{[0,-1,1]}^{-1}+E_{[0,-1,-1]}^{-1}\right)+\frac{1}{\lambda\sqrt{2}}\ub_2 e^{-\phi_2+\eta}\left(E_{[0,1,-1]}^{-1}-E_{[0,1,1]}^{-1}\right)
\eea
where $u_i, \phi_i, \Phi_i, \pa \beta_{ij}, H, \nu$ are all arbitrary functions of $z,\zb$.  Nothing special is meant by partial derivatives in front of any functions: these are just names of arbitrary functions, and where an integrated form appears, this is an arbitrary, but fixed integral.  The bars over the functions also do not denote anything special, this simply denotes different functions.  Note that the above is in fact an arbitrary set of functions.  For example, the coefficients where $\eta$ appears (without a derivative) are controlled completely by the set of functions $\phi_2, u_i$. The dressing by $\eta$ is simply a convenience: any integral of the coefficient of $H_2$ in $A$ would do.  Finally, we also denote $\beta_{ij}=-\beta_{ji}$ (similarly for $\betab$) for later convenience.  Further, the indices $i,j$ appearing on $\beta$ and $u$ are to be raised and lowered by
\be
|\eta_{ij}|=\begin{pmatrix} -1 & 0 & 0 \\ 0 & 1 & 0 \\ 0 & 0 & 1 \end{pmatrix}.
\ee
(we apologize for the notation: we will always include $\eta_{ij}$ as the matrix above, and $\eta$ without indices to denote a field).

Next, let us pin down some of the above coefficients.  Note that the matrix structure $(E_{[0,-1,1]}^{-1}+E_{[0,-1,-1]}^{-1})\lambda^{-1}$ only appears in $\pa \Ab_{-1}$, and in $[A_0,\Ab_{-1}]$.  The term $[A_0,\Ab_{-1}]$ has some parts determined by the Cartan subalgebra, and others that come from $E_{[i,j,k]}$ commutators. However, $(E_{[0,-1,1]}^{-1}+E_{[0,-1,-1]}^{-1})\lambda^{-1}$ differs from all the other terms in $\Ab_{-1}$ by having a $-1$ entry as it's second lower index: all others have $+1$.  Hence, none of these other matrices can be transformed into $(E_{[0,-1,1]}^0+E_{[0,-1,-1]}^0)\lambda^{-1}$ by a single commutator.  Using this, we find the simple equation
\be
\Phi_2=\pa \phi_2.
\ee
Similarly, we find
\be
\Phib_2=\pab \phib_2
\ee
Given this, we may actually make a gauge transformation of the form $e^{H_2 \frac12(\phi_2-\phib_2)}$ to make all occurrences of the fields $\phi_2$ and $\phib_2$ occur in the combination $\phi_2+\phib_2$.  Another way of saying this is that the equations of motion arising from the above Lax pair do not depend on $\phi_2-\phib_2$, and so we may gauge this component to zero (using $e^{H_2 \frac12(\phi_2-\phib_2)}$), and so enforces $\phi_2-\phib_2=0$.  Therefore, without loss of generality
\be
\phib_2=\phi_2.
\ee
Likewise, we may use a gauge transformation of the form $e^{\D \frac12(\eta-\etab)}$ and $e^{\kh\frac12(\nu-\nub)}$ to remove these differences and give
\be
\etab=\eta, \qquad \nub=\nu.
\ee

We may also make some reality cuts on the above connection (this is some $Z_2$ action).  For this, we note that
\bea
&& \pab A - \pa \Ab + [A,\Ab]=0 \nn \\
\rightarrow_{\uln{\;}} && \pab \uln A - \pa \uln \Ab + [\uln A, \uln \Ab]=0 \nn \\
\rightarrow_{*} && \pa (\uln A^*) - \pab (\uln \Ab^*) + [\uln A^*,\uln \Ab^*] =0 \nn \\
\rightarrow_{O} && \pa ((\uln A^*)^O) - \pab ((\Ab^*)^O) + [(\uln A^*)^O, (\uln \Ab^*)^O] \nn \\
&&= -\left(\pab ((\uln \Ab^*)^O)-\pa ((\uln A^*)^O)+[(\uln \Ab^*)^O,(\uln A^*)^O]\right)=0 \nn \\
\eea
Hence, if $A,\Ab$ define a flat connection, so do $ ((\uln \Ab^*)^O, (\uln A^*)^O)$.  We therefore require that
\be
A(\lambda)=\left(\uln \Ab\left(\frac{1}{\lambda^*}\right)^*\right)^O, \qquad \Ab(\lambda)=\left(\uln A\left(\frac{1}{\lambda^*}\right)^*\right)^O
\ee
This boils down to the requirement that
\be
\phi_i^* = \bar \phi_i, \qquad (\pa \beta_{ij})^*=\pab\betab_{ij}, \qquad \ub_i=u_i^*.
\ee

Now let us work on the coefficients $E_{[-1,1,0]}^{-1}\lambda^{-1}$.  These again appear only in very specific combinations, and we find the equation
\bea
&&-\pa\left(e^{-\phi_2 +\eta}(\ub_0+\ub_1)\right)+\left(\left(\pa \beta_{01}-\pa \phi_2+\pa\eta\right)e^{-\phi_2 +\eta}(\ub_0+\ub_1)\right)\nn \\
&& \qquad \qquad \qquad \qquad \qquad \qquad+(\pa\beta_{02}+\pa\beta_{12})\ub_2e^{-\phi_2+\eta}=0 \nn \\
&&\rightarrow \pa \beta_{01} (\ub_0+\ub_1)- \pa(\ub_0+\ub_1)+(\pa\beta_{02}+\pa\beta_{12})\ub_2 \nn \\
&&=   -(\pa \ub_0- \pa\beta_{0} \!^2 \ub_2-\pa \beta_{0}\!^1\ub_1)-(\pa \ub_1- \pa\beta_{1}\! ^2 \ub_2-\pa\beta_{1} \!^0 \ub_0) =0 \label{E0p1}
\eea
for $E_{[1,1,0]}^{-1}\lambda^{-1}$, we find
\bea
&&-\pa\left(e^{-\phi_2 +\eta}(\ub_0-\ub_1)\right)+\left(\left(-\pa \beta_{01}-\pa \phi_2+\pa\eta\right)e^{-\phi_2 +\eta}(\ub_0-\ub_1)\right) \nn \\
&&\qquad \qquad \qquad \qquad \qquad \qquad+(\pa\beta_{02}-\pa\beta_{12})\ub_2e^{-\phi_2+\eta}=0 \nn \\
&& \rightarrow -\pa \beta_{01} (\ub_0-\ub_1)- \pa(\ub_0-\ub_1)+(\pa\beta_{02}-\pa\beta_{12})\ub_2 \nn \\
&&= -(\pa \ub_0- \pa\beta_{0} \!^2 \ub_2-\pa \beta_{0}\!^1\ub_1)+(\pa \ub_1- \pa\beta_{1}\! ^2 \ub_2-\pa\beta_{1} \!^0 \ub_0) =0 \label{E0m1}
\eea
and finally for the matrix structure $\left(E_{[0,1,-1]}^{-1}-E_{[0,1,1]}^{-1}\right)\lambda^{-1}$ we find
\bea
&&-\pa\left(\ub_2e^{-\phi_2+\eta}\right)+\left(-\pa\phi_2+\pa \eta \right)\ub_2e^{-\phi_2+\eta} +\frac12\left(\pa\beta_{02}+\pa\beta_{12}\right)(\ub_0-\ub_1)e^{-\phi_2+\eta} \nn \\
&&\qquad \qquad \qquad \qquad \qquad \qquad +\frac12\left(\pa\beta_{02}-\pa\beta_{12}\right)(\ub_0+\ub_1)e^{-\phi_2+\eta}=0 \nn \\
&&\rightarrow -\pa u_2 +\pa\beta_{02}\ub_0-\pa\beta_{12}\ub_1 \nn \\
&&= -\pa u_2 +\pa \beta_{2}\!^0 \ub_0 + \pa \beta_{2}\!^1\ub_1=0 \label{E2}
\eea
The equations (\ref{E0p1}),(\ref{E0m1}), (\ref{E2}) may be written in a compact form
\be
\pa \ub_i = \pa\beta_i\,^j \ub_j.\label{ubdiffeq}
\ee
We recall that $\pa\beta_{ij}=-\pa\beta_{ji}$ and $ij$ are raised and lowered by $\eta_{ij}$ so that $\beta^{ij}$ is antisymmetric as well.  This gives that
\be
\frac12 \pa \left(\eta^{ij} \ub_i \ub_j\right) = \ub_i \pa\beta^{ij}\ub_j=0
\ee
so that
\be
\eta^{i j} \ub_i \ub_j=\Ub(\zb)^2 \label{antiholofunction}
\ee
for some antiholomorphic function $\Ub(\zb)$.

We may similarly deduce (or simply use the reality cut above) to show that
\be
\pab u_i = \pab\betab_i \,^j u_j. \label{udiffeq}
\ee
and
\be
\frac12 \pa \left(\eta^{ij} u_i u_j\right) = u_i \pa\beta^{ij}u_j=0
\ee
and so
\be
\eta^{i j} u_i u_j=U(z)^2 \label{holofunction}
\ee
for some holomorphic function $U(z)$.

As shown in our previous work \cite{Burrington:2009bh}, the pair of equations (\ref{holofunction}), (\ref{udiffeq}) is locally equivalent to the equations
\be
u_i = U(z) R_i\,^j v_j, \quad \pab \betab_i \,^j= (\pab R R^{-1})_i\,^j \label{ubetabslns}
\ee
where $v_0=1$ and $v_1=v_2=0$, and similarly (\ref{antiholofunction}), (\ref{ubdiffeq}) are locally equivalent to the equations
\be
\ub_i = \Ub(\zb) \Rb_i\,^j v_j, \quad \pa \beta_i \,^j= (\pa \Rb \Rb^{-1})_i\,^j. \label{ubbetaslns}
\ee
where $R$ and $\bar{R}$ are general $SO(1,2:{\mathbb C})$ matrix.  The reality cut now simply reads that
\be
\bar{R}=R^*, \qquad \bar{\phi}_2=\phi_2^*.
\ee

We note that one may have to be careful about the above identification (\ref{ubbetaslns}), because $\pab \betab_i \,^j= (\pab R R^{-1})_i\,^j+(R\delta V R^{-1})_i\,^j$ is also a solution, with $\delta V_i\,^j$ satisfying $\delta V_i\,^j v_j=0$.  It can be shown that $\delta V_1\,^2$ is the only non trivial term: this is therefore some unconstrained $U(1)$ connection that may have non trivial ``Wilson line'' components around some special points, but may locally be removed.  We henceforth take that (\ref{ubetabslns}) and (\ref{ubbetaslns}) are globally satisfied.  In any case, all of the $\lambda^1$ and $\lambda^{-1}$ equations of motion have been satisfied.

Further, now the discussion exactly parallels that of \cite{Burrington:2009bh}: the equations of motion coming from the $H_i^0, E_{[i,j,k]}^0$, written in terms of $R$ and $\Rb$, are identical to that of \cite{Burrington:2009bh}, with an additional factor of $e^{2\eta}$ multiplying the ``potential terms.''  This, however, does not change any of the covariance properties of left multiplying $R$ by a real $SO(1,2)$ matrix.  Further, the equation of motion coming from $\D$ is always $\pa\pab \eta=0$, and the equation of motion coming from $\kh$ only couples to $u_0 \ub_0-u_1 \ub_1-u_2 \ub_2$, and so are invariant under this left multiplication as well.  Therefore, this then allows for an identical treatment, removing degrees of freedom that don't couple to the equations similar to \cite{Burrington:2009bh}.  This procedure will result in giving exactly the Lax connection (\ref{Epnew})-(\ref{connectionnew}).

\section{A related one dimensional integrable model.}

\label{dimreduct}

We now turn to the question of how to generate solutions for the equations via reducing to a 1D integrable model.  Particularly important for us is when $U(z)^2=p(z)$ for some polynomial $z$.  Now we remind the reader that the presence of $U(z)$ in the equations of motion (and in the action) can be absorbed into a redefinition of $\eta$ and $\phi_2$.  This redefinition is $\eta= \eta' - \frac14 \ln(U(z) \bar{U}(\zb))$, and $\phi_2=\phi_2'+\frac14 \ln(U(z) \bar{U}(\zb))$.  Since $U$ is now related to a polynomial, we find that the redefinition becomes $\eta= \eta' + \frac18 \sum_i \ln(|(z-z_i)|^2)$ where $z_i$ are the locations of the zeros of the polynomial $p$.  This has the simple effect that it adds delta functions to the right hand side of the equations of motion.  Therefore, we may view all models with different $U$ as being the same model, simply with different delta function sources for $\eta$ and $\phi_2$.  The model of concern is therefore the conformally invariant model with $U$ set to $1$.

Further, we note that every solution to the conformally invariant model descends to a solution of a model with a fixed $U$ by conformally gauge fixing $\eta$ to be $\frac14\ln\left(U \Ub\right)$, and further that any solution of the original model (with $U$ given) can be lifted to a solution of the conformal model.  We will now explore the conformally invariant model.

Above we have argued that the different solutions are specified by different delta function sources, and so we will begin by trying to look at the model where there is only one delta function source.  While this is not very interesting by itself, it would be interesting to know how or if one can use this type of solution as a basic building block to generate the solutions of interest.  First, note that a single delta function source is rotationally symmetric around this point.  Hence there is a rotationally symmetric solution to this problem.  We find it first convenient to change variables from $z={\mathcal C}^{-2-n} (z')^{-n}$.  For this rotationally invariant solution, we find it convenient to map the plane to the cylinder by $z=e^{w}$, and we write $w=x+iy$.  The rotationally symmetric solutions are then given by solutions where all fields depend only on $x$.  Thus, the conformally invariant action
\bea
{\mathcal{L}}=&&\pa \phi_1 \pab \phi_1 + \pa \phi_2 \pab \phi_2 + \tanh^2(\phi_1)\pa \Lambda_1 \pab \Lambda_1 + \frac12 \left(e^{-2\phi_2}\cosh(2\phi_1)+e^{2\phi_2}\right)e^{2\eta} \nn \\
&&+ (\pa \nu \pab \eta + \pab \nu \pa \eta).
\eea
is reduced to the one dimensional action
\bea
{\mathcal{L}}_1=&&\pa_x\phi_1  \pa_x \phi_1 + \pa_x \phi_2 \pa_x \phi_2 + \tanh^2(\phi_1)\pa_x \Lambda_1 \pa_x \Lambda_1 + \frac12\left(e^{-2\phi_2}\cosh(2\phi_1)+e^{2\phi_2}\right)e^{2\eta} \nn \\
&&+ 2(\pa_x \nu \pa_x \eta).
\eea
The Lax pair is easy to write down for the above system, given the Lax pair for the original system.  Note that if the original system satisfied $\pa_x A_y- \pa_y A_x + [A_y, A_x]$, then when all fields are $y$ independent, we have that $\pa_x A_y = [A_y, -A_x]$.  This just identifies the Lax pair as $L=A_y$ and $M=-A_x$ with $L'=[L,M]$.  However, recall that in this case the ``matrices'' above are defined as members of an affine Lie algebra, so some care is needed to interpret the results.

Now we segue to find an easier system to analyze.  If we take that the value of $\eta=0$ above, the system reduces further, and we may neglect the coupling to $\nu$ and $\eta$ altogether, and so ignore the affine extension of the algebra.  This special case admits the Lax pair
\bea
L&=& i\Bigg(-\pa_x \phi_1 H_1 -\pa_x \phi_2 H_2 \nn \\
&& -\frac12 \frac{\pa_x \Lambda_1 \sinh(\phi_1)}{\cosh^2(\phi_1)}\left(E_{[1,0,-1]}+E_{[1,0,1]}+E_{[-1,0,1]}+E_{[1,0,-1]}\right) \nn \\
&& +\frac{\lambda}{2\sqrt{2}}\left(e^{\phi_1-\phi_2}E_{[1,-1,0]}+ e^{-\phi_1-\phi_2}E_{[-1,-1,0]}+e^{\phi_2}\left(E_{[0,1,-1]}+E_{[0,1,1]}\right)\right) \\
&& -\frac{1}{2\sqrt{2}\lambda }\left(e^{\phi_1-\phi_2}E_{[-1,1,0]}+ e^{-\phi_1-\phi_2}E_{[1,1,0]}+e^{\phi_2}\left(E_{[0,-1,1]}+E_{[0,-1,-1]}\right)\right)\Bigg)
\eea
and
\bea
 M &=& -\frac12 \frac{\pa_x \Lambda_1}{\cosh(\phi_1)}\left(E_{[1,0,-1]}+E_{[1,0,1]}-E_{[-1,0,1]}-E_{[1,0,-1]}\right) \nn \\
&& -\frac{\lambda}{2\sqrt{2}}\left(e^{\phi_1-\phi_2}E_{[1,-1,0]}+ e^{-\phi_1-\phi_2}E_{[-1,-1,0]}+e^{\phi_2}\left(E_{[0,1,-1]}+E_{[0,1,1]}\right)\right) \\
&& -\frac{1}{ 2\sqrt{2}\lambda}\left(e^{\phi_1-\phi_2}E_{[-1,1,0]}+ e^{-\phi_1-\phi_2}E_{[1,1,0]}+e^{\phi_2}\left(E_{[0,-1,1]}+E_{[0,-1,-1]}\right)\right).
\eea
From this we may read the conserved quantities from $\Tr(L^n)$.  In this case, there are 3 conserved quantities, namely
\bea
&& H=(\pa_x \phi_1)^2 + (\pa_x \phi_2)^2  + \tanh^2(\phi_1)(\pa_x \Lambda_1)^2 - \frac12 \left(e^{-2\phi_2}\cosh(2\phi_1)+e^{2\phi_2}\right), \\
&& P=\tanh^2(\phi_1)\pa_x \Lambda_1 \\
&& Q=\frac{4\sinh^2(\phi_1) (\pa_x \Lambda_1)^2 (\pa_x \phi_2)^2}{\cosh^4(\phi_1)}- \frac{2\sinh^2(\phi_1)\left(\sinh^2(\phi_1)e^{-2\phi_2}+e^{2\phi_2}\right)(\pa_x \Lambda_1)^2}{\cosh^4(\phi_1)} \nn \\
&&\quad +4(\pa_x \phi_1)^2(\pa_x \phi_2)^2+4\sinh(\phi_1)\cosh(\phi_1) e^{-2\phi_2}(\pa_x \phi_1)(\pa_x\phi_2) \nn \\
&& \quad -2e^{2\phi_2}(\pa_x \phi_1)^2+\left(\frac12 e^{-2\phi_2}\cosh(2\phi_1)+\frac12 e^{2\phi_2}\right)^2 -\frac14 \cosh(4\phi_2).
\eea
We have not explicitly checked this type of Ansatz for a solution, however, it would be quite interesting to see if this generates any new classes of minimal surfaces in AdS$_5$.


\begin{thebibliography}{99}

\bibitem{Maldacena:1997re}
  J.~M.~Maldacena,
  Adv.\ Theor.\ Math.\ Phys.\  {\bf 2}, 231 (1998)
  [Int.\ J.\ Theor.\ Phys.\  {\bf 38}, 1113 (1999)]
  [arXiv:hep-th/9711200].
  E.~Witten,
  Adv.\ Theor.\ Math.\ Phys.\  {\bf 2}, 253 (1998)
  [arXiv:hep-th/9802150].

\bibitem{Kachru:1998ys}
  S.~Kachru and E.~Silverstein,
  Phys.\ Rev.\ Lett.\  {\bf 80}, 4855 (1998)
  [arXiv:hep-th/9802183].
  I.~R.~Klebanov and E.~Witten,
  Nucl.\ Phys.\  B {\bf 536}, 199 (1998)
  [arXiv:hep-th/9807080].

\bibitem{Maldacena:1998im}
  J.~M.~Maldacena,
  Phys.\ Rev.\ Lett.\  {\bf 80}, 4859 (1998)
  [arXiv:hep-th/9803002].
  S.~J.~Rey and J.~T.~Yee,
  Eur.\ Phys.\ J.\  C {\bf 22}, 379 (2001)
  [arXiv:hep-th/9803001].

\bibitem{Drukker:2000rr}
  N.~Drukker and D.~J.~Gross,
  J.\ Math.\ Phys.\  {\bf 42}, 2896 (2001)
  [arXiv:hep-th/0010274].
  N.~Drukker, D.~J.~Gross and H.~Ooguri,
  Phys.\ Rev.\  D {\bf 60}, 125006 (1999)
  [arXiv:hep-th/9904191].

\bibitem{Berenstein:1998ij}
  D.~E.~Berenstein, R.~Corrado, W.~Fischler and J.~M.~Maldacena,
  Phys.\ Rev.\  D {\bf 59}, 105023 (1999)
  [arXiv:hep-th/9809188].

\bibitem{Zarembo:1999bu}
  K.~Zarembo,
  Phys.\ Lett.\  B {\bf 459}, 527 (1999)
  [arXiv:hep-th/9904149].
  P.~Olesen and K.~Zarembo,
  arXiv:hep-th/0009210.
  B.~A.~Burrington and L.~A.~P.~Zayas,
  arXiv:1012.1525 [hep-th].
  L.~F.~Alday and A.~A.~Tseytlin,
  arXiv:1105.1537 [hep-th].

\bibitem{Drukker:2005cu}
  N.~Drukker and B.~Fiol,
  JHEP {\bf 0601}, 056 (2006)
  [arXiv:hep-th/0506058].


\bibitem{Gross:1998gk}
  D.~J.~Gross and H.~Ooguri,
  Phys.\ Rev.\  D {\bf 58}, 106002 (1998)
  [arXiv:hep-th/9805129].

\bibitem{Kruczenski:2002fb}
  M.~Kruczenski,
  JHEP {\bf 0212}, 024 (2002)
  [arXiv:hep-th/0210115].


\bibitem{Kruczenski:2007cy}
  M.~Kruczenski, R.~Roiban, A.~Tirziu and A.~A.~Tseytlin,
  Nucl.\ Phys.\  B {\bf 791}, 93 (2008)
  [arXiv:0707.4254 [hep-th]].



\bibitem{classicalStringSpin}
  S.~S.~Gubser, I.~R.~Klebanov and A.~M.~Polyakov,
  Nucl.\ Phys.\  B {\bf 636}, 99 (2002)
  [arXiv:hep-th/0204051].
  S.~Frolov and A.~A.~Tseytlin,
  JHEP {\bf 0206}, 007 (2002)
  [arXiv:hep-th/0204226].
  A.~A.~Tseytlin,
  arXiv:hep-th/0311139.
  A.~A.~Tseytlin,
  arXiv:hep-th/0409296.


\bibitem{Frolov:2003qc}
  S.~Frolov and A.~A.~Tseytlin,
  Nucl.\ Phys.\  B {\bf 668}, 77 (2003)
  [arXiv:hep-th/0304255].




\bibitem{Arutyunov:2003za}
  G.~Arutyunov, J.~Russo and A.~A.~Tseytlin,
  Phys.\ Rev.\  D {\bf 69}, 086009 (2004)
  [arXiv:hep-th/0311004].




\bibitem{Arutyunov:2003uj}
  G.~Arutyunov, S.~Frolov, J.~Russo and A.~A.~Tseytlin,
  Nucl.\ Phys.\  B {\bf 671}, 3 (2003)
  [arXiv:hep-th/0307191].

\bibitem{Janik:2010gc}
  R.~A.~Janik, P.~Surowka, A.~Wereszczynski,
  JHEP {\bf 1005}, 030 (2010).
  [arXiv:1002.4613 [hep-th]].


\bibitem{Buchbinder:2010vw}
  E.~I.~Buchbinder and A.~A.~Tseytlin,
  JHEP {\bf 1008}, 057 (2010)
  [arXiv:1005.4516 [hep-th]].


\bibitem{Roiban:2010fe}
  R.~Roiban and A.~A.~Tseytlin,
  Phys.\ Rev.\  D {\bf 82}, 106011 (2010)
  [arXiv:1008.4921 [hep-th]].


\bibitem{Ryang:2010bn}
  S.~Ryang,
  JHEP {\bf 1101}, 092 (2011)
  [arXiv:1011.3573 [hep-th]].

\bibitem{Hernandez:2010tg}
  R.~Hernandez,
  J.\ Phys.\ A  {\bf 44}, 085403 (2011)
  [arXiv:1011.0408 [hep-th]].



\bibitem{GluonScat}
  L.~F.~Alday and J.~M.~Maldacena,
  JHEP {\bf 0706}, 064 (2007)
  [arXiv:0705.0303 [hep-th]].
  L.~F.~Alday,
  Fortsch.\ Phys.\  {\bf 56}, 816 (2008)
  [arXiv:0804.0951 [hep-th]].
  L.~F.~Alday and R.~Roiban,
  Phys.\ Rept.\  {\bf 468}, 153 (2008)
  [arXiv:0807.1889 [hep-th]].


\bibitem{Pohlmeyer:1975nb}
  K.~Pohlmeyer,
  Commun.\ Math.\ Phys.\  {\bf 46}, 207 (1976).


\bibitem{Pohlmeyer:1975nb2}
   H.~J.~De Vega and N.~G.~Sanchez,
  Phys.\ Rev.\  D {\bf 47}, 3394 (1993).
   M.~Grigoriev and A.~A.~Tseytlin,
  Nucl.\ Phys.\  B {\bf 800}, 450 (2008)
  [arXiv:0711.0155 [hep-th]].
  R.~Roiban and A.~A.~Tseytlin,
  JHEP {\bf 0904}, 078 (2009)
  [arXiv:0902.2489 [hep-th]].
  B.~Hoare, Y.~Iwashita and A.~A.~Tseytlin,
  J.\ Phys.\ A  {\bf 42}, 375204 (2009)
  [arXiv:0906.3800 [hep-th]].
  J.~L.~Miramontes,
  JHEP {\bf 0810}, 087 (2008)
  [arXiv:0808.3365 [hep-th]].
  T.~J.~Hollowood and J.~L.~Miramontes,
  JHEP {\bf 0904}, 060 (2009)
  [arXiv:0902.2405 [hep-th]].




\bibitem{Burrington:2009bh}
  B.~A.~Burrington and P.~Gao,
  JHEP {\bf 1004}, 060 (2010)
  [arXiv:0911.4551 [hep-th]].

\bibitem{Alday:2009dv}
  L.~F.~Alday, D.~Gaiotto and J.~Maldacena,
  arXiv:0911.4708 [hep-th].

\bibitem{Beisert:2010jr}
  N.~Beisert {\it et al.},
  arXiv:1012.3982 [hep-th].

\bibitem{Jevicki:2007aa}
  A.~Jevicki, K.~Jin, C.~Kalousios and A.~Volovich,
  JHEP {\bf 0803}, 032 (2008)
  [arXiv:0712.1193 [hep-th]].

\bibitem{Alday:2009yn}
  L.~F.~Alday and J.~Maldacena,
  JHEP {\bf 0911}, 082 (2009)
  [arXiv:0904.0663 [hep-th]].



\bibitem{Babelon:1990bq}
  O.~Babelon and L.~Bonora,
  Phys.\ Lett.\  B {\bf 244}, 220 (1990).


\bibitem{Constantinidis:1992hs}
  C.~P.~Constantinidis, L.~A.~Ferreira, J.~F.~Gomes and A.~H.~Zimerman,
  Phys.\ Lett.\  B {\bf 298}, 88 (1993)
  [arXiv:hep-th/9207061].


\bibitem{Kac:1990gs}
  V.~G.~Kac,
{\it  Cambridge, UK: Univ. Pr. (1990) 400 p}

\bibitem{DiFrancesco:1997nk}
  P.~Di Francesco, P.~Mathieu and D.~Senechal,
{\it  New York, USA: Springer (1997) 890 p}

\bibitem{CATstudies}
  A.~Fring, G.~Mussardo and P.~Simonetti,
  Nucl.\ Phys.\  B {\bf 393}, 413 (1993)
  [arXiv:hep-th/9211053].
  H.~Aratyn, C.~P.~Constantinidis, L.~A.~Ferreira, J.~F.~Gomes and A.~H.~Zimerman,
  Nucl.\ Phys.\  B {\bf 406}, 727 (1993)
  [arXiv:hep-th/9212086].
  D.~I.~Olive, N.~Turok and J.~W.~R.~Underwood,
  Nucl.\ Phys.\  B {\bf 401}, 663 (1993).
 H.~Aratyn, C.~P.~Constantinidis, L.~A.~Ferreira, J.~F.~Gomes and A.~H.~Zimerman,
  arXiv:hep-th/9304080.
  H.~Aratyn, L.~A.~Ferreira, J.~F.~Gomes and A.~H.~Zimerman,
  Mod.\ Phys.\ Lett.\  A {\bf 9}, 2783 (1994)
  [arXiv:hep-th/9308086].
  G.~Papadopoulos and B.~J.~Spence,
  Mod.\ Phys.\ Lett.\  A {\bf 9}, 1579 (1994)
  [arXiv:hep-th/9402079].
  B.~Y.~Hou, B.~Y.~Hou, X.~H.~Wang, C.~H.~Xiong and R.~H.~Yue,
  arXiv:hep-th/0406250.

\bibitem{Aratyn:1990tr}
  H.~Aratyn, L.~A.~Ferreira, J.~F.~Gomes and A.~H.~Zimerman,
  Phys.\ Lett.\  B {\bf 254}, 372 (1991).

\bibitem{Leznov:1979td}
  A.~N.~Leznov and M.~V.~Savelev,
  Lett.\ Math.\ Phys.\  {\bf 3}, 489 (1979).

\bibitem{Babelon}
  O.~Babelon, D.~Bernard and M.~Talon
{\it  Cambridge, UK: Univ. Pr. (2003) 602 p}

\bibitem{Ishizeki:2011bf}
  R.~Ishizeki, M.~Kruczenski and S.~Ziama,
  arXiv:1104.3567 [hep-th].

\bibitem{Babich:1992mc}
  M.~Babich and A.~Bobenko,
   Duke Mathematical Journal {\bf 72} , {\bf No. 1}, 151 (1993).









\end{thebibliography}
\end{document}